\documentclass[a4paper,12pt]{amsart}
\usepackage{amsmath}
   \usepackage{bm}
\usepackage{amsthm,amscd}
\usepackage{multirow}
\usepackage{mathrsfs}
\usepackage{verbatim}
\usepackage{amsfonts,amsmath,latexsym,amssymb,txfonts,bm}
\usepackage[american]{babel}
\usepackage{dsfont}
\usepackage{braket}
\newcommand\bea{\begin{eqnarray}}
\newcommand\eea{\end{eqnarray}}

\newcommand{\diff}{\mathrm{d}}
\newcommand{\cal}{\mathcal}

\begin{document}

\title{Spectral Functions for Regular Sturm-Liouville Problems}

\author{Guglielmo Fucci}\thanks{GF is partially supported by the East Carolina University Thomas Harriot College of Arts and Sciences Summer research grant and the ORAU Ralph E. Powe Junior Faculty Enhancement Award.}
\address{Department of Mathematics, East Carolina University,
Greenville, NC 27858, USA}
\email{fuccig@ecu.edu}
\author{Curtis Graham}
\address{Department of Mathematics, Baylor University, Waco, TX 76798, USA}
\email{Curtis\textunderscore Graham@baylor.edu}
\author{Klaus Kirsten}
\address{Department of Mathematics, Baylor University, Waco, TX 76798, USA}
\email{Klaus\textunderscore Kirsten@baylor.edu}

\begin{abstract}
In this paper we provide a detailed analysis of the analytic continuation of the spectral zeta
function associated with one-dimensional regular Sturm-Liouville problems endowed with self-adjoint
separated and coupled boundary conditions. The spectral zeta function is represented in terms of a
complex integral and the analytic continuation in the entire complex plane is achieved by using the
Liouville-Green (or WKB) asymptotic expansion of the eigenfunctions associated with the problem. The
analytically continued expression of the spectral zeta function is then used to compute the functional
determinant of the Sturm-Liouville operator and the coefficients of the asymptotic expansion of the
associated heat kernel.
\end{abstract}
\date{\today}
\maketitle

\section{Introduction}

The analysis of the spectrum of Laplace-type operators acting on suitable functions defined on a Riemannian manifold
is a subject of extreme importance in several areas of both mathematics and physics. In quantum field theory, for instance, this
spectrum represents the energy content of the modes that constitute a quantum field. Despite its importance the exact
computation of the spectrum is not possible in general with the exception of very few special cases. This lack of exact knowledge
of the spectrum, however, does not constitute a major problem since very often one is more interested in obtaining
information that strictly depends on the entire spectral sequence rather than on the explicit value of each individual
term of the spectrum. This type of global information about the spectrum of a Laplace-type operator is encoded in the
spectral functions and can be obtained from their analysis. The most widely used spectral functions in mathematics and
physics are the spectral zeta function and the heat kernel \cite{elizalde94,gilkey95,kirsten01,vassilevich03}.

For Laplace-type operators, $\Delta$, defined on a compact Riemannian manifold with or without boundary, the spectrum
$\{l_{n}\}_{n\in\mathbb{N}}$ is discrete, bounded from below, and forms an increasing sequence of real numbers with a
unique accumulation point at infinity \cite{gilkey95}. The spectral zeta function is then defined in terms of the eigenvalues as follows
\begin{equation}\label{0}
  \zeta(s)=\sum_{n=1}^{\infty}l_{n}^{-s}\;,
\end{equation}
where $s$ is a complex variable and each eigenvalue is counted with its finite multiplicity. If we denote with $D$ the
dimension of the Riemannian manifold under consideration, then, for $n\to\infty$, the eigenvalues behave according to Weyl's
estimate as $l_{n}\sim n^{2/D}$ \cite{gilkey95}. This asymptotic behavior implies that the spectral zeta function defined above
converges in the region $\Re(s)>D/2$. However, it was proved in \cite{mina49} that $\zeta(s)$ can be analytically continued to
a meromorphic function in the entire complex $s$-plane which develops only simple poles and is holomorphic at the point $s=0$ .

The heat kernel $\theta(t)$ is related to the spectral zeta function by Mellin transform \cite{gilkey95,voros87}
and is defined as the trace of the heat semigroup $\exp\{-t\Delta\}$ in the Hilbert space of square integrable
functions on the Riemannian manifold. The function $\theta(t)$ is well defined for $\Re(t)>0$ and its explicit
expression in terms of the eigenvalues is
\begin{equation}
  \theta(t)=\sum_{n=1}^{\infty}e^{-tl_{n}}\;.
\end{equation}
Due to the fact that in general the eigenvalues are not known, the heat kernel cannot be computed explicitly.
However, the asymptotic expansion of $\theta(t)$ for $t\to 0$ can be found in general situations and its coefficients
are expressed only in terms of geometric invariants of the underlying manifold \cite{gilkey75,gilkey04,kirsten01,vassilevich03}.

The two spectral functions defined above constitute the main mathematical tool used to regularize divergent quantities
in the ambit of quantum field theory on Minkowski or Riemannian manifolds \cite{birrell,fulling,hawking77}. For instance,
the spectral zeta function is used to compute the functional determinant of an elliptic operator \cite{ray} which is
proportional to the one-loop effective action. Moreover, zeta function techniques are of fundamental importance in the
study of the Casimir energy of a quantum system \cite{blau88,bord01-353-1,bord09b}. The coefficients of the small-$t$
asymptotic expansion of the heat kernel explicitly identify the infrared divergences in the one-loop effective action.
Such terms are extremely relevant in the process of renormalizing a theory of quantum fields \cite{birrell}.

Sturm-Liouville problems appear frequently in many areas of mathematics and physics and for this reason they represent
one of the most studied subjects in the theory of differential equations. They arise naturally, for instance, in the
study of quantum fields propagating in a Riemannian manifold possessing particular symmetries. In fact, in quantum
field theory the dynamics of a quantum field is described by a Laplace operator defined on a $D$-dimensional Riemannian
manifold with or without boundary. If the manifold under consideration has symmetries that at every point allow for the existence
of a metric tensor of the form
\begin{equation}
  g_{\mu\nu}=\begin{pmatrix} 1 & 0 \\ 0 & \eta_{ij} \end{pmatrix}\;,
\end{equation}
with $i,j=\{2,\ldots,D\}$, then the Laplace operator acting on scalar fields $\vartheta$,
\begin{equation}
   \Delta\vartheta=-\frac{1}{\sqrt{\det g}}\partial_{\mu}\left(g^{\mu\nu}\sqrt{\det g}\;\partial_{\nu}\right)\vartheta\; ,
\end{equation}
can be expressed, assuming that $\partial_{x_{1}}\eta_{ij}=0$, as
\begin{equation}
  \Delta\vartheta=\left(-\frac{\diff^{2}}{\diff x_{1}^{2}}+\Delta_{\eta}\right)\vartheta\;,
\end{equation}
where $\Delta_{\eta}$ denotes the Laplacian associated with the coordinates $\{x_{2},\ldots, x_{D}\}$.
In this situation the eigenvalue equation $\Delta\vartheta=\lambda^{2}\vartheta$, which describes the energy content of the
quantum field, can be solved by separation of variables $\vartheta=\theta(x_{1})\omega(x_{2},\ldots, x_{D})$. By denoting
with $\nu^{2}$ the eigenvalues of $\Delta_{\eta}$ and, in addition, by assuming that the field propagates under the influence of
external conditions described by a smooth potential $V(x_{1})$, the eigenvalues $\lambda$ are obtained from the one-dimensional differential equation
\begin{equation}
  \left(-\frac{\diff^{2}}{\diff x_{1}^{2}}+V(x_{1})+\nu^{2}\right)\theta(x_{1})=\lambda^{2}\theta(x_{1})\;,
\end{equation}
with suitable self-adjoint boundary conditions imposed on $\theta(x_{1})$. This type of differential equation endowed with self-adjoint
boundary conditions represents a particular example of the more general class of regular Sturm-Liouville problems.

The main focus of this paper is on the analysis of the spectral zeta function and the heat kernel asymptotic expansion for regular one-dimensional
Sturm-Liouville problems endowed with general self-adjoint boundary conditions. Due to the generality of this problem, standard techniques used to
perform the analytic continuation of the spectral zeta function, $\zeta(s)$, become unsuitable in this situation. In fact, in the majority of cases
the spectral zeta function and its analytic continuation are computed when either the eigenvalues or the eigenfunctions of the problem are explicitly known.
For general one-dimensional Sturm-Liouville problems, however, this information is not available and, hence, other approaches for the analysis of $\zeta(s)$
have to be devised. One method for performing the analytic continuation of the spectral zeta function that does not require any explicit knowledge of
either the eigenvalues or the eigenfunctions has been developed in \cite{fucci1,fucci2} and relies only on the WKB analysis of the eigenvalue problem.
This method is applicable to a wide variety of eigenvalue equations and boundary conditions and it will be employed in this work to perform
the analytic continuation of $\zeta(s)$ associated with regular one-dimensional Sturm-Liouville problems.

The boundary conditions imposed on a system play an important role in the study of its associated spectral functions.
In the vast majority of the literature the spectral functions are obtained for specific types of differential
operators and for particular boundary conditions. This is somewhat undesirable since for a given problem it is
often necessary to perform a separate analysis for each type of boundary condition imposed. Unlike other methods,
the one we develop and utilize in this work will provide a more comprehensive approach since it is suitable for
treating the most general one-dimensional second-order differential operator endowed with \emph{any} self-adjoint
boundary condition. To our knowledge this work represents the first effort to develop a technique suitable for
the analysis of spectral functions associated with a symmetric second-order differential operator in one dimension
endowed with any self-adjoint boundary condition.

The outline of the paper is as follows. In the next section we will describe the one-dimensional Sturm-Liouville
problem that will be the object of the subsequent analysis and the associated self-adjoint boundary conditions. In Section
\ref{sec3} we present the WKB analysis of the one-dimensional Sturm-Liouville operator with separated or coupled boundary conditions.
Section \ref{sec4} focuses on the analytic continuation of the spectral zeta function which is based on the results of the WKB analysis.
In Section \ref{sec5} we describe the modifications to the approach that are necessary when zero modes are present.
In Sections \ref{sec6} and \ref{sec7} we utilize the analytically continued expression of the spectral zeta function to compute the zeta
regularized functional determinant of the Sturm-Liouville operator and the coefficients of the heat kernel asymptotic expansion, respectively.

 \section{Regular Sturm-Liouville Problems}\label{sec2}

In this work we will be mainly concerned with the most general Sturm-Liouville differential operator. This symmetric second order differential operator,
which we denote by $\cal{L}$, has the form \cite{zettl}
\begin{equation}\label{1}
{\cal L}=-\frac{\diff}{\diff x}\left(p(x)\frac{\diff}{\diff x}\right)+V(x)\;,
\end{equation}
and acts on scalar functions defined on the interval $I=[0,1]\subset\mathbb{R}$. This choice of interval does not lack in generality since it is
clear that the linear transformation $x^{\prime}=a+(b-a)x$ with $a,b\in\mathbb{R}$ yields an operator of the same form as (\ref{1}) but defined
on the more general one-dimensional interval $[a,b]$. In addition we assume that $p(x)>0$ for $x\in I$ and that both $p(x)$ and $V(x)$ belong to
${\mathcal C}^\infty (I)$.

For the symmetric operator $\mathcal{L}$ in (\ref{1}) we consider the following differential equation
\begin{equation}\label{2}
{\mathcal L}\,\varphi_{\lambda}(x)=\lambda^{2}\varphi_{\lambda}(x)\;,
\end{equation}
where $\lambda\in\mathbb{C}$ and $\varphi_{\lambda}(x)\in C^{2}(I)$. Spectral functions associated with differential equations of the form (\ref{2}) containing a differential
operator of the type given in (\ref{1}) can be conveniently studied by using the first-order formalism \cite{kirsten03,kirsten04}. By defining the vector
\begin{equation}\label{3}
\mathrm{Y}_{\lambda}(x)=\begin{pmatrix}
 \varphi_{\lambda}(x) \\
    p(x)\varphi^{\prime}_{\lambda}(x) \\
\end{pmatrix}\;,
\end{equation}
with the prime denoting, here and in the rest of the paper, differentiation with respect to the variable $x$, we can rewrite (\ref{2}) as a
system of first-order differential equations
\begin{equation}\label{4}
\frac{\diff}{\diff x}\mathrm{Y}_{\lambda}(x)=\mathrm{A}_{\lambda}(x)\mathrm{Y}_{\lambda}(x)\;,
\end{equation}
with the $2\times 2$ matrix $\mathrm{A}_{\lambda}(x)$ defined as
\begin{equation}\label{5}
\mathrm{A}_{\lambda}(x)=\begin{pmatrix}
0 & p^{-1}(x) \\
V(x)-\lambda^{2} & 0
\end{pmatrix}\;.
\end{equation}

Amongst all the solutions to the differential equation (\ref{2}), or equivalently to the system (\ref{4}), we only select the ones
that satisfy self-adjoint boundary conditions. With this restriction placed on the solutions, the numbers $\lambda\in\mathbb{R}$ become the
eigenvalues and (\ref{2}) coupled with self-adjoint boundary conditions represents a regular Sturm-Liouville problem. The self-adjoint
boundary conditions that can be imposed on the solutions to (\ref{2}) can be naturally divided into two classes \cite{zettl}.
	
\emph{Separated boundary conditions} have the following general form
\begin{eqnarray}\label{6}
A_{1}\varphi_{\lambda}(0)-A_{2}p(0)\varphi^{\prime}_{\lambda}(0)&=&0\;,\nonumber\\
B_{1}\varphi_{\lambda}(1)+B_{2}p(1)\varphi^{\prime}_{\lambda}(1)&=&0\;,
\end{eqnarray}
with $A_{1},A_{2},B_{1},B_{2}\in\mathbb{R}$ and $(A_{1},A_{2})\neq(0,0)$, and $(B_{1},B_{2})\neq (0,0)$. In the first-order formalism
these boundary conditions can be written in matrix form as
\begin{eqnarray}\label{7}
\begin{pmatrix}
A_{1} & -A_{2} \\
0 & 0
\end{pmatrix}\begin{pmatrix}
\varphi_{\lambda}(0) \\
p(0)\varphi^{\prime}_{\lambda}(0)
\end{pmatrix}+\begin{pmatrix}
0 & 0 \\
B_{1} & B_{2} \\
\end{pmatrix}\begin{pmatrix}
\varphi_{\lambda}(1) \\
p(1)\varphi^{\prime}_{\lambda}(1)
\end{pmatrix}=\begin{pmatrix}
0 \\
0
\end{pmatrix}\;.
\end{eqnarray}
We would like to point out that when $A_{2}=B_{2}=0$ the equations (\ref{6}) reduce to the Dirichlet boundary conditions, while for
$A_2 \neq 0$, $B_2 \neq 0$, the system (\ref{6})
represents Robin boundary conditions; for details regarding this identification see Section 7. 
Mixed or hybrid boundary conditions are obtained by setting $A_{1}=B_{2}=0$ or $A_{2}=B_{1}=0$.

\emph{Coupled boundary conditions} can be expressed in general as
\begin{equation}\label{8}
\begin{pmatrix}
\varphi_{\lambda}(1) \\
p(1)\varphi^{\prime}_{\lambda}(1)
\end{pmatrix}=e^{i\gamma}\mathrm{K}\begin{pmatrix}
\varphi_{\lambda}(0) \\
p(0)\varphi^{\prime}_{\lambda}(0)
\end{pmatrix}\;,
\end{equation}
where $-\pi<\gamma\leq 0$ or $0\leq\gamma<\pi$ and $\mathrm{K}\in\textrm{SL}_{2}(\mathbb{R})$. Note that when $\gamma=0$ and
$\mathrm{K}=I_{2}$, with $I_{2}$ representing the $2\times 2$ identity matrix, we obtain the familiar periodic boundary conditions.

Due to the fact that the functions $p(x)$ and $V(x)$ are arbitrary elements of ${\mathcal C}^\infty (I)$, 
the eigenvalues and corresponding eigenfunctions of (\ref{2}) endowed with either separated or coupled boundary conditions cannot
be found explicitly. However, by imposing either of the previously described self-adjoint boundary conditions to a general solution
of the equation (\ref{2}) we will be able to obtain an expression whose zeroes implicitly determine the eigenvalues.

In more detail, let $\varphi_{\lambda}(x)$ denote a general solution to the equation (\ref{2}). For each $\lambda\in\mathbb{C}$
we choose $\varphi_{\lambda}(x)$ and $p(x)\varphi^{\prime}_{\lambda}(x)$ such that they satisfy the following \emph{initial conditions}
\begin{equation}\label{9}
\varphi_{\lambda}(0)=A_{2}\;,\quad\textrm{and}\quad p(0)\varphi^{\prime}_{\lambda}(0)=A_{1}\;.
\end{equation}
With this particular choice of initial conditions, the first equation of the boundary conditions (\ref{6}) is automatically
satisfied while the second equation, describing the boundary condition at the endpoint $x=1$, provides an implicit equation
for the eigenvalues $\lambda$, namely
\begin{equation}\label{10}
\Omega(\lambda)=B_{1}\varphi_{\lambda}(1)+B_{2}p(1)\varphi^{\prime}_{\lambda}(1)=0\;.
\end{equation}
Since the boundary conditions (\ref{6}) are self-adjoint, all the zeroes of the equation (\ref{10}) are real and positive \cite{zettl}.

For coupled boundary conditions, instead, we write the general solution $\varphi_{\lambda}(x)$ to (\ref{2}) as a linear combination of two
linearly independent solutions $u_{\lambda}(x)$ and $v_{\lambda}(x)$
\begin{equation}\label{11}
\varphi_{\lambda}(x)=\alpha u_{\lambda}(x)+\beta v_{\lambda}(x)\;,
\end{equation}
with arbitrary coefficients $\alpha$ and $\beta$. For each $\lambda\in\mathbb{C}$ the functions $u_{\lambda}(x)$ and $v_{\lambda}(x)$
are uniquely determined as solutions to (\ref{2}) satisfying the initial conditions \cite{bai96}
\begin{equation}\label{12}
u_{\lambda}(0)=0\;,\quad\textrm{and}\quad p(0)u^{\prime}_{\lambda}(0)=1\;,
\end{equation}
for $u_{\lambda}(x)$, and
\begin{equation}\label{13}
v_{\lambda}(0)=1\;,\quad\textrm{and}\quad p(0)v^{\prime}_{\lambda}(0)=0\;,
\end{equation}
for $v_{\lambda}(x)$. The coupled boundary conditions in (\ref{8}) together with the formula for $\varphi_{\lambda}(x)$ in
(\ref{11}) lead to the following homogeneous linear system in the unknowns $\alpha$ and $\beta$
\begin{eqnarray}\label{14}
\alpha\left[u_{\lambda}(1)-e^{i\gamma}k_{12}\right]+\beta\left[v_{\lambda}(1)-e^{i\gamma}k_{11}\right]&=&0\;,\nonumber\\
\alpha\left[p(1)u^{\prime}_{\lambda}(1)-e^{i\gamma}k_{22}\right]+\beta\left[p(1)v^{\prime}_{\lambda}(1)-e^{i\gamma}k_{21}\right]&=&0\;,
\end{eqnarray}
where we have denoted with $k_{ij}$ the entries of the matrix $\mathrm{K}$ and we have used the initial conditions
$\varphi_{\lambda}(0)=\beta$ and $p(0)\varphi^{\prime}_{\lambda}(0)=\alpha$ which follow directly from the relations (\ref{12}) and (\ref{13}).
In order for the system (\ref{14}) to have a non-trivial solution the determinant of its coefficient matrix must vanish.
This condition provides the following implicit equation for the eigenvalues $\lambda$
\begin{equation}\label{15}
\Delta(\lambda)=\left(u_{\lambda}(1)-e^{i\gamma}k_{12}\right)\left(p(1)v^{\prime}_{\lambda}(1)-e^{i\gamma}k_{21}\right)-\left(v_{\lambda}(1)-e^{i\gamma}k_{11}\right)\left(p(1)u^{\prime}_{\lambda}(1)-e^{i\gamma}k_{22}\right)=0\;.
\end{equation}
The above implicit equation, however, can be simplified further. In fact, by performing the products and by using the
relation $\textrm{det}\mathrm{K}=1$, which follows from the fact that $\mathrm{K}\in\textrm{SL}_{2}(\mathbb{R})$, we obtain
\begin{eqnarray}\label{16}
\Delta(\lambda)&=&u_{\lambda}(1)p(1)v^{\prime}_{\lambda}(1)-v_{\lambda}(1)p(1)u^{\prime}_{\lambda}(1)-e^{2i\gamma}\nonumber\\
&+&e^{i\gamma}\left[k_{22}v_{\lambda}(1)-k_{21}u_{\lambda}(1)+k_{11}p(1)u_{\lambda}^{\prime}(1)-k_{12}p(1)v_{\lambda}^{\prime}(1)\right]=0\;.
\end{eqnarray}
The first two terms on the right hand side of (\ref{16}) constitute the Wronskian $W[u_{\lambda}(x),v_{\lambda}(x)]$ of the two
linearly independent solutions $u_{\lambda}(x)$ and $v_{\lambda}(x)$ computed at $x=1$. Since $W^{\prime}[u_{\lambda}(x),v_{\lambda}(x)]=0$
we can conclude that $W[u_{\lambda}(x),v_{\lambda}(x)]=C$. To find the constant $C$ we set $x=0$ and use the initial conditions (\ref{12})
and (\ref{13}) to obtain $C=W[u_{\lambda}(0),v_{\lambda}(0)]=-\textrm{det}\,I_{2}=-1$. The last remark proves that
\begin{equation}\label{17}
u_{\lambda}(1)p(1)v^{\prime}_{\lambda}(1)-v_{\lambda}(1)p(1)u^{\prime}_{\lambda}(1)=-1\;,
\end{equation}
and, consequently, the equation (\ref{16}) which implicitly determines the eigenvalues $\lambda$ becomes \cite{bai96}
\begin{equation}\label{18}
\Delta(\lambda)=2\cos\gamma-\left[k_{22}v_{\lambda}(1)-k_{21}u_{\lambda}(1)+k_{11}p(1)u_{\lambda}^{\prime}(1)-k_{12}p(1)v_{\lambda}^{\prime}(1)\right]=0\;.
\end{equation}

The spectral zeta function associated with the eigenvalue equation (\ref{2}) endowed with separated or coupled boundary conditions is
\begin{equation}\label{19}
\zeta(s)=\sum_{\lambda}\lambda^{-2s}\;,
\end{equation}
which is convergent, due to Weyl's estimate \cite{gilkey95}, in the semi-plane $\Re(s)>1/2$. In order to perform the analytic continuation
we represent $\zeta(s)$ in (\ref{19}) in terms of a contour integral by exploiting Cauchy's residue theorem.
In the case of separated boundary conditions the eigenvalues are implicitly given by (\ref{10}) and, hence, the corresponding spectral
zeta function can be expressed as \cite{kirsten01}
\begin{equation}\label{20}
\zeta^{\mathrm{S}}(s)=\frac{1}{2\pi i}\int_{\mathcal{C}}\diff\lambda\,\lambda^{-2s}\frac{\partial}{\partial\lambda}\ln\Omega(\lambda)\;.
\end{equation}
Similarly, for coupled boundary conditions the spectral zeta function can be represented as
\begin{equation}\label{21}
\zeta^{\mathrm{C}}(s)=\frac{1}{2\pi i}\int_{\mathcal{D}}\diff\lambda\,\lambda^{-2s}\frac{\partial}{\partial\lambda}\ln\Delta(\lambda)\;,
\end{equation}
with the function $\Delta(\lambda)$ given by (\ref{18}).
In the above integral representations $\mathcal{C}$ and $\mathcal{D}$ are contours in the complex plane that encircle in the
counterclockwise direction all the roots, which are on the positive real axis, of $\Omega(\lambda)$ and $\Delta(\lambda)$, respectively.
The complex integrals in (\ref{20}) and (\ref{21}) are well defined, by construction, in the region $\Re(s)>1/2$.

By deforming the contour of integration $\mathcal{C}$ in (\ref{20}) to the imaginary axis and by using the property that
$\Omega(i\lambda)=\Omega(-i\lambda)$, which can be proved from (\ref{10})
by noticing that $\varphi_{i\lambda}(x)=\varphi_{-i\lambda}(x)$ and $\varphi^{\prime}_{i\lambda}(x)=\varphi^{\prime}_{-i\lambda}(x)$, we obtain,
by changing variables $\lambda\to iz$, the representation
\begin{equation}\label{22}
\zeta^{\mathrm{S}}(s)=\frac{\sin\pi s}{\pi}\int_{0}^{\infty}\diff z\,z^{-2s}\frac{\partial}{\partial z}\ln\Omega(iz)\;.
\end{equation}
By following a procedure analogous to the one used to obtain (\ref{22}) from (\ref{20}), one can prove that for coupled boundary conditions the spectral zeta function takes the form
\begin{equation}\label{23}
\zeta^{\mathrm{C}}(s)=\frac{\sin\pi s}{\pi}\int_{0}^{\infty}\diff z\,z^{-2s}\frac{\partial}{\partial z}\ln\Delta(iz)\;.
\end{equation}
From the differential equation (\ref{2}) one can show that the leading $z\to\infty$ behavior of its solutions
is exponential growth, see (\ref{31}). This remark and the explicit expressions
(\ref{10}) and (\ref{18}) allow us to conclude that the functions $\Omega(iz)$ and $\Delta(iz)$ have a similar exponential
behavior as $z\to\infty$ and, hence, the integral representations (\ref{22}) and (\ref{23}) converge at the upper limit of
integration for $\Re(s)>1/2$. Furthermore, since the solutions of the differential equation (\ref{2}) are analytic with respect to the
parameter $\lambda$, the functions $\Omega(iz)$ and $\Delta(iz)$ are analytic in the variable $z$.
In particular this implies that in a neighborhood of $z=0$ we have that $\Omega(iz)-\Omega(0)\sim z^{2}$ and
linear terms in $z$ are not present since $\Omega(iz)=\Omega(-iz)$.
By assuming that $z=0$ is not an eigenvalue of our problem, we can conclude that for $z\to 0$ we have $\ln\Omega(iz)= \ln \Omega (0) + {\cal O} (z^{2})$.
This last estimate implies that the integral representation (\ref{22}) converges at the lower limit of integration for $\Re(s)<1$.
A completely similar analysis can be performed for the function $\Delta(iz)$ leading to the conclusion that the integral representation
(\ref{23}) is also convergent at the lower limit of integration for $\Re(s)<1$. We can therefore state that the integral
representations (\ref{22}) and (\ref{23}) converge in the region $1/2<\Re(s)<1$.

\section{The WKB analysis of the Sturm-Liouville problem}\label{sec3}
To perform the analytic continuation of $\zeta^{\mathrm{S}}(s)$ and $\zeta^{\mathrm{C}}(s)$ to values lying outside of the
strip $1/2<\Re(s)<1$ we will rely on the asymptotic expansion for large $z$ of the eigenfunctions of the Sturm-Liouville
problem (\ref{2}) endowed with separated or coupled boundary conditions \cite{kirsten01}. To start the analysis we consider
the differential equation (\ref{2}), with $\lambda=iz$,
\begin{equation}\label{24}
\left[-\frac{\diff}{\diff x}\left(p(x)\frac{\diff}{\diff x}\right)+V(x)\right]\varphi_{iz}(x)=-z^{2}\varphi_{iz}(x)\;.
\end{equation}
By utilizing the following ansatz for the solution
\begin{equation}\label{25}
\varphi_{iz}(x)=\exp\left\{\int_{0}^{x}\mathcal{S}(t,z)\diff t\right\}\;,
\end{equation}
and by substituting it into the differential equation (\ref{24}), we obtain a first order nonlinear differential equation for $\mathcal{S}(x,z)$
\begin{equation}\label{26}
\left[p(x)\mathcal{S}(x,z)\right]^{\prime}=V(x)+z^{2}-p(x)\mathcal{S}^{2}(x,z)\;.
\end{equation}
The last differential equation, although nonlinear, is very suitable for the application of WKB techniques to obtain the
large-$z$ behavior of $\mathcal{S}(x,z)$ and consequently, through the relation (\ref{25}), of $\varphi_{iz}(x)$  \cite{bend10b}.

For $z\to\infty$ we assume that $\mathcal{S}(x,z)$ has an asymptotic expansion of the form
\begin{equation}\label{27}
\mathcal{S}(x,z)\sim zS_{-1}(x)+\sum_{i=0}^{\infty}\frac{S_{i}(x)}{z^{i}}\;.
\end{equation}
By substituting the above expansion into the differential equation (\ref{26}) and by equating like powers of $z$ we obtain
\begin{equation}\label{28}
S_{-1}^{\pm}(x)=\pm\frac{1}{\sqrt{p(x)}}\;,\qquad S_{0}^{\pm}(x)=-\frac{1}{2}\frac{\diff}{\diff x}\ln\left(p(x)S_{-1}^{\pm}(x)\right)\;,
\end{equation}
for the leading and the first subleading term of the asymptotic expansion. For the term with $i=1$ we have
\begin{equation}\label{29}
S_{1}^{\pm}(x)=\frac{1}{2p(x)S_{-1}^{\pm}(x)}\left[V(x)-p(x)\left(S_{0}^{\pm}\right)^{2}(x)-\left(p(x)S_{0}^{\pm}(x)\right)^{\prime}\right]\;,
\end{equation}
while for the higher asymptotic orders with $i\geq 1$ we obtain the recurrence relation
\begin{equation}\label{30}
S^{\pm}_{i+1}(x)=-\frac{1}{2p(x)S_{-1}^{\pm}(x)}\left[\left(p(x)S_{i}^{\pm}(x)\right)^{\prime}+p(x)\sum_{m=0}^{i}S_{m}^{\pm}(x)S_{i-m}^{\pm}(x)\right]\;.
\end{equation}
The $\pm$ represents the different choice of sign in the leading term $S_{-1}$ of the asymptotic expansion (\ref{27}).
The different signs correspond to the two solutions $\mathcal{S}^{+}(z,x)$ and $\mathcal{S}^{-}(z,x)$ to the differential equation (\ref{26}).
These solutions provide, through the relation (\ref{25}), the exponentially growing and decaying parts of
the function $\varphi_{iz}(x)$. The correct asymptotic behavior of $\varphi_{iz}(x)$ for $z\to\infty$ is obtained
as a linear combination of the asymptotically
increasing and decaying terms as follows
\begin{equation}\label{31}
\varphi_{iz}(x)=A\exp\left\{\int_{0}^{x}\mathcal{S}^{+}(t,z)\diff t\right\}+B\exp\left\{\int_{0}^{x}\mathcal{S}^{-}(t,z)\diff t\right\}\;,
\end{equation}
where the terms $A$ and $B$ are uniquely determined once the initial conditions are imposed.
\subsection{Separated Boundary Conditions}
For separated boundary conditions we impose the initial conditions (\ref{9}) on the function $\varphi_{iz}(x)$ in (\ref{31}) to obtain
\begin{equation}\label{32}
  A=-\frac{A_{2}p(0)\mathcal{S}^{-}(0,z)-A_{1}}{p(0)\left[\mathcal{S}^{+}(0,z)-\mathcal{S}^{-}(0,z)\right]}\;,\qquad
  B=\frac{A_{2}p(0)\mathcal{S}^{+}(0,z)-A_{1}}{p(0)\left[\mathcal{S}^{+}(0,z)-\mathcal{S}^{-}(0,z)\right]}\;.
\end{equation}
By substituting the expressions (\ref{32}) into (\ref{31}) and then by using the resulting expression for $\varphi_{iz}(x)$ in the formula for $\Omega(iz)$ in (\ref{10}) we obtain
\begin{equation}\label{33}
  \Omega(iz)=-\frac{\left[A_{2}p(0)\mathcal{S}^{-}(0,z)-A_{1}\right]\left[B_{2}p(1)\mathcal{S}^{+}(1,z)+B_{1}\right]}{p(0)\left[\mathcal{S}^{+}(0,z)-\mathcal{S}^{-}(0,z)\right]}
  \exp\left\{\int_{0}^{1}\mathcal{S}^{+}(t,z)\diff t\right\}\left(1+\mathcal{E}(z)\right)\;,
\end{equation}
where we have denoted with $\mathcal{E}(z)$ exponentially small contributions in $z$. The integral representation
of the spectral zeta function $\zeta^{\mathrm{S}}(s)$ in (\ref{22}) contains the term $\ln\Omega(iz)$ and, therefore, its asymptotic expansion is
needed for the process of analytic continuation of $\zeta^{\mathrm{S}}(s)$. From the expression (\ref{33}) we have
\begin{eqnarray}\label{34}
  \ln\Omega(iz)&=&\ln\left[-A_{2}p(0)\mathcal{S}^{-}(0,z)+A_{1}\right]+\ln\left[B_{2}p(1)\mathcal{S}^{+}(1,z)+B_{1}\right]
  \nonumber\\
  &-&\ln\left[p(0)\left(\mathcal{S}^{+}(0,z)-\mathcal{S}^{-}(0,z)\right)\right]+\int_{0}^{1}\mathcal{S}^{+}(t,z)\diff t+\tilde{\mathcal{E}}(z)\;,
\end{eqnarray}
where $\tilde{\mathcal{E}}(z)$ denotes exponentially decaying terms. At this point, it is necessary to further
expand for $z\to\infty$ each term appearing in (\ref{34}). From the results (\ref{28})-(\ref{30}) one obtains
\begin{equation}\label{35}
  \mathcal{S}^{+}(0,z)-\mathcal{S}^{-}(0,z)=\frac{2z}{\sqrt{p(0)}}\left[1+\sum_{i=1}^{\infty}\frac{\omega_{i}(0)}{z^{i+1}}\right]\;,
\end{equation}
where for $i\in\mathbb{N}^{+}$
\begin{equation}\label{36}
  \omega_{i}(0)=\frac{\sqrt{p(0)}}{2}\left[S_{i}^{+}(0)-S_{i}^{-}(0)\right]\;.
\end{equation}
Since one can prove, from the recurrence relation (\ref{30}), that $S_{i}^{+}$ and $S_{i}^{-}$ satisfy the relation
\begin{equation}\label{37}
  S_{i}^{-}(x)=(-1)^{i}S_{i}^{+}(x)\;,
\end{equation}
we have, from (\ref{36}), that
\begin{equation}\label{38}
  \omega_{2i}(0)=0\;,\qquad \omega_{2i+1}(0)=\sqrt{p(0)}S_{2i+1}^{+}(0)\;.
\end{equation}
We use the results (\ref{35}), (\ref{36}) and (\ref{38}) to obtain
\begin{equation}\label{39}
  \ln\left[p(0)\left(\mathcal{S}^{+}(0,z)-\mathcal{S}^{-}(0,z)\right)\right]=\frac{1}{2}\ln p(0)+\ln 2z+\sum_{i=1}^{\infty}\frac{D_{2i-1}(0)}{z^{2i}}\;,
\end{equation}
where the terms $D_{2i-1}(0)$ are determined through the cumulant expansion
\begin{equation}\label{40}
  \ln\left[1+\sum_{k=0}^{\infty}\frac{\omega_{2k+1}(0)}{z^{2k+2}}\right]\simeq \sum_{i=1}^{\infty}\frac{D_{2i-1}(0)}{z^{2i}}\;.
\end{equation}

More care is needed in the expansion of the first two terms on the right hand side of (\ref{34}). For these terms the cases $A_{2}=B_{2}=0$, and $A_{2}=0$ or $B_{2}=0$ need to be distinguished from all the other cases.
This distinction is necessary since the large $z$ asymptotic behavior of $\ln\Omega(iz)$ critically depends on whether $A_{2}$ or $B_{2}$, or both, vanish.
In the case $A_{2}=B_{2}=0$, which can be easily recognized to represent Dirichlet boundary
conditions, the arguments of $\ln\left[-A_{2}p(0)\mathcal{S}^{-}(0,z)+A_{1}\right]$ and $\ln\left[B_{2}p(1)\mathcal{S}^{+}(1,z)+B_{1}\right]$
reduce to just constants and, hence, no further analysis is necessary. In order to avoid analyzing
each case separately, we introduce the function
\begin{equation}
\delta(x)=\left\{\begin{array}{ll}
1 & \textrm{if}\; x=0\\
0 & \textrm{if}\; x\neq 0
\end{array}\right.\;,
\end{equation}
and replace the first term on the right hand side of (\ref{34}) by
\begin{equation}\label{40a}
\left[1-\delta(A_{2})\right]\ln\left[-A_{2}p(0)\mathcal{S}^{-}(0,z)+A_{1}\right]+\delta(A_{2})\ln A_{1}\;,
\end{equation}
and the second term by
\begin{equation}\label{40b}
\left[1-\delta(B_{2})\right]\ln\left[B_{2}p(1)\mathcal{S}^{+}(1,z)+B_{1}\right]+\delta(B_{2})\ln B_{1}\;.
\end{equation}

From the asymptotic expansion (\ref{27}), we obtain
\begin{equation}\label{41}
 -A_{2}p(0)\mathcal{S}^{-}(0,z)+A_{1}=A_{2}\sqrt{p(0)}\,z +\sum_{i=0}^{\infty}\frac{\sigma_{i}^{-}(0)}{z^{i}}\;,
\end{equation}
and
\begin{equation}\label{42}
B_{2}p(1)\mathcal{S}^{+}(1,z)+B_{1}=B_{2}\sqrt{p(1)}\,z+\sum_{i=0}^{\infty}\frac{\sigma_{i}^{+}(1)}{z^{i}}\;,
\end{equation}
where, by recalling (\ref{37}),
\begin{eqnarray}\label{43}
  \sigma_{0}^{-}(0)=-A_{2}p(0)S_{0}^{+}(0)+A_{1}\;,\qquad \sigma_{i}^{-}(0)=(-1)^{i+1}A_{2}p(0)S_{i}^{+}(0)\;,\quad i\geq 1
\end{eqnarray}
and
\begin{equation}\label{44}
  \sigma_{0}^{+}(1)=B_{2}p(1)S_{0}^{+}(1)+B_{1}\;,\qquad \sigma_{i}^{+}(1)=B_{2}p(1)S_{i}^{+}(1)\;,\quad i\geq 1\;.
\end{equation}
The results obtained in (\ref{41}) and (\ref{42}) lead to the expansions
\begin{eqnarray}\label{45}
\ln\left[-A_{2}p(0)\mathcal{S}^{-}(0,z)+A_{1}\right]=\ln \left(A_{2}\sqrt{p(0)}\right)+\ln z+\sum_{i=1}^{\infty}\frac{\mathcal{Z}_{i}^{-}(0)}{z^{i}}\;,
\end{eqnarray}
and
\begin{eqnarray}\label{46}
  \ln\left[B_{2}p(1)\mathcal{S}^{+}(1,z)+B_{1}\right]=\ln \left(B_{2}\sqrt{p(1)}\right)+\ln z+\sum_{i=1}^{\infty}\frac{\mathcal{Z}_{i}^{+}(1)}{z^{i}}\;,
\end{eqnarray}
where $\mathcal{Z}_{i}^{-}(0)$ and $\mathcal{Z}_{i}^{+}(1)$ are found through the relations
\begin{eqnarray}
  \ln\left[1+\frac{1}{A_{2}\sqrt{p(0)}}\sum_{i=1}^{\infty}\frac{\sigma_{i-1}^{-}(0)}{z^{i}}\right]&\simeq& \sum_{k=1}^{\infty}\frac{\mathcal{Z}_{k}^{-}(0)}{z^{k}}\;,\label{47a}\\
  \ln\left[1+\frac{1}{B_{2}\sqrt{p(1)}}\sum_{i=1}^{\infty}\frac{\sigma_{i-1}^{+}(1)}{z^{i}}\right]&\simeq& \sum_{k=1}^{\infty}\frac{\mathcal{Z}_{k}^{+}(1)}{z^{k}}\;.\label{47b}
\end{eqnarray}
By using the expansions (\ref{27}) and (\ref{39}) and by recalling that for a unified treatment of all separated boundary conditions we need to replace the first two terms of the right hand side of (\ref{34}) by (\ref{40a}) and (\ref{40b}), whose expansions can be obtained from (\ref{41}) and (\ref{42}), we find
\begin{eqnarray}\label{48}
  \ln\Omega(iz)&=&-\frac{1}{4}\ln \left( p(0)p(1) \right)+\left[1-\delta(A_{2})\right]\ln \left( A_{2}\sqrt{p(0)}\right)
  +\left[1-\delta(B_{2})\right]\ln \left( B_{2}\sqrt{p(1)}\right) +\delta(A_{2})\ln A_{1}\nonumber\\
&+&\delta(B_{2})\ln B_{1} -\ln 2z+\left[2-\delta(A_{2})-\delta(B_{2})\right]\ln z+z\int_{0}^{1}S_{-1}^{+}(t)\diff t+\sum_{i=1}^{\infty}\frac{\mathcal{M}_{i}}{z^{i}}\;,
\end{eqnarray}
where we have discarded exponentially decreasing terms and we have used the relation
\begin{equation}\label{50}
  \int_{0}^{1}S_{0}^{+}(t)\diff t=-\frac{1}{4}\ln\frac{p(1)}{p(0)}\;.
\end{equation}
Since, for $i\in\mathbb{N}^{+}$, $D_{2i}(0)=0$, we have that, when $i=2m+1$ with $m\in\mathbb{N}_{0}$,
\begin{equation}\label{51}
\mathcal{M}_{2m+1}=\int_{0}^{1}S_{2m+1}^{+}(t)\diff t+\left[1-\delta(A_{2})\right]\mathcal{Z}^{-}_{2m+1}(0)+\left[1-\delta(B_{2})\right]\mathcal{Z}^{+}_{2m+1}(1)\;,
\end{equation}
while, for $i=2m$ with $m\in\mathbb{N}^{+}$,
\begin{equation}\label{52}
  \mathcal{M}_{2m}=\int_{0}^{1}S_{2m}^{+}(t)\diff t-D_{2m-1}(0)+\left[1-\delta(A_{2})\right]\mathcal{Z}^{-}_{2m}(0)+\left[1-\delta(B_{2})\right]\mathcal{Z}^{+}_{2m}(1)\;.
\end{equation}

\subsection{Coupled Boundary Conditions}

The implicit equation for the eigenvalues in the case of coupled boundary conditions, namely $\Delta(iz)=0$ in (\ref{18}), contains
two linearly independent solutions $u_{iz}(x)$ and $v_{iz}(x)$ to (\ref{24}) which are found by imposing the
initial conditions (\ref{12}) and (\ref{13}) to a general solution of (\ref{24}). For $z\to\infty$ both functions $u_{iz}(x)$ and $v_{iz}(x)$
can be expressed as a linear combination of an exponentially increasing and an exponentially decreasing term
in the same way as indicated in (\ref{31}).
By imposing the initial conditions (\ref{12}) to the general solution in (\ref{31}) we obtain for $A$ and $B$ the relations
\begin{eqnarray}\label{54}
  A+B=0\;,\quad A=\frac{1}{p(0)\left[\mathcal{S}^{+}(0,z)-\mathcal{S}^{-}(0,z)\right]}\;,
\end{eqnarray}
which lead to the following result
\begin{equation}\label{55}
  u_{iz}(x)=\frac{1}{p(0)\left[\mathcal{S}^{+}(0,z)-\mathcal{S}^{-}(0,z)\right]}\left[\exp\left\{\int_{0}^{x}\mathcal{S}^{+}(t,z)\diff t\right\}-\exp\left\{\int_{0}^{x}\mathcal{S}^{-}(t,z)\diff t\right\}\right]\;.
\end{equation}
For the function $v_{iz}(x)$ we impose, instead, the initial conditions (\ref{13}) to (\ref{31}) to obtain
\begin{equation}\label{56}
  A+B=1\;,\quad A=-\frac{\mathcal{S}^{-}(0,z)}{\mathcal{S}^{+}(0,z)-\mathcal{S}^{-}(0,z)}\;,
\end{equation}
which provide the following expression
\begin{equation}\label{57}
  v_{iz}(x)=\frac{1}{\mathcal{S}^{+}(0,z)-\mathcal{S}^{-}(0,z)}\left[-\mathcal{S}^{-}(0,z)\exp\left\{\int_{0}^{x}\mathcal{S}^{+}(t,z)\diff t\right\}+\mathcal{S}^{+}(0,z)\exp\left\{\int_{0}^{x}\mathcal{S}^{-}(t,z)\diff t\right\}\right]\;.
\end{equation}
Obviously, expressions for the functions $u^{\prime}_{iz}(x)$ and $v^{\prime}_{iz}(x)$, which appear in $\Delta(iz)$,
are obtained by simply differentiating (\ref{55}) and (\ref{57}). The last remark and the explicit expressions (\ref{55}) and (\ref{57}) can be used in
the formula for $\Delta(iz)$ in (\ref{18}) to obtain
\begin{eqnarray}\label{58}
  \Delta(iz)&=&\frac{\left[-k_{21}-k_{22}p(0)\mathcal{S}^{-}(0,z)+k_{11}p(1)\mathcal{S}^{+}(1,z)
  +k_{12}p(1)p(0)\mathcal{S}^{-}(0,z)\mathcal{S}^{+}(1,z)\right]}{p(0)\left[\mathcal{S}^{+}(0,z)-\mathcal{S}^{-}(0,z)\right]}\nonumber\\
  &\times&\exp\left\{\int_{0}^{1}\mathcal{S}^{+}(t,z)\diff t\right\}\left(1+\varepsilon(z)\right)\; ,
\end{eqnarray}
where exponentially small contributions have been collectively denoted by $\varepsilon(z)$. From this expression it is
not very difficult to obtain
\begin{eqnarray}\label{59}
  \ln\Delta(iz)&=&-\ln\left[p(0)\left(\mathcal{S}^{+}(0,z)-\mathcal{S}^{-}(0,z)\right)\right]+\int_{0}^{1}\mathcal{S}^{+}(t,z)\diff t\nonumber\\
  &+&\ln\left[-k_{21}-k_{22}p(0)\mathcal{S}^{-}(0,z)+k_{11}p(1)\mathcal{S}^{+}(1,z)
  +k_{12}p(1)p(0)\mathcal{S}^{-}(0,z)\mathcal{S}^{+}(1,z)\right]+\tilde{\varepsilon}(z)\;,
\end{eqnarray}
with $\tilde{\varepsilon}(z)$ being exponentially decreasing terms as $z\to\infty$. The expression (\ref{59}) is the starting point for
the computation of the large-$z$ asymptotic expansion of $\ln\Delta(iz)$. The asymptotic expansion of the first two terms on the right hand side of
(\ref{59}) has already been found in the case of separated boundary conditions and, hence, will not be repeated here.
Instead, we will concentrate only on the last logarithmic term in (\ref{59}). From the asymptotic
expansion of $\mathcal{S}^{+}$ and $\mathcal{S}^{-}$ we obtain, for the terms linear in $\mathcal{S}^{+}$ and $\mathcal{S}^{-}$
\begin{eqnarray}\label{60}
  -k_{21}-k_{22}p(0)\mathcal{S}^{-}(0,z)+k_{11}p(1)\mathcal{S}^{+}(1,z)=z\left(k_{22}\sqrt{p(0)}+k_{11}\sqrt{p(1)}\right)+\sum_{i=0}^{\infty}\frac{\Phi_{i}}{z^{i}}\;,
\end{eqnarray}
with
\begin{eqnarray}\label{61}
  \Phi_{0}& = &-k_{21}-k_{22}p(0)S_{0}^{+}(0)+k_{11}p(1)S_{0}^{+}(1)\;,\\
  \Phi_{i}&=&(-1)^{i+1}k_{22}p(0)S_{i}^{+}(0)+k_{11}p(1)S_{i}^{+}(1)\;,\quad i\geq 1\;, \nonumber
\end{eqnarray}
where in the previous relations we have exploited (\ref{37}). In addition, for the term proportional to $\mathcal{S}^{-}\mathcal{S}^{+}$  we have
\begin{equation}\label{62}
k_{12}p(1)p(0)\mathcal{S}^{-}(0,z)\mathcal{S}^{+}(1,z)=-z^{2}k_{12}\sqrt{p(0)p(1)}\left[1+\sum_{i=1}^{\infty}z^{-i}\left(\sum_{m=0}^{i}(-1)^{m}\bar{S}_{m}(0)\bar{S}_{i-m}(1)\right)\right]\;,
\end{equation}
with $\bar{S}_{0}(x)=1$ and, for $i\in\mathbb{N}^{+}$,
\begin{equation}\label{63}
  \bar{S}_{i}(x)=\sqrt{p(x)}S_{i-1}^{+}(x)\;.
\end{equation}
In (\ref{62}) we have used, once again, the relation (\ref{37}) in order to express $S_{i}^{-}(x)$ in terms of $S_{i}^{+}(x)$.

From the relations (\ref{60}) and (\ref{62}) one can notice that the leading term of the asymptotic expansion
of the argument of the last logarithmic term in (\ref{59}) strictly depends on whether or not the coefficient
$k_{12}$ of the matrix $\mathrm{K}$ vanishes. Because of this different leading behavior as $z\to\infty$ we
will need to distinguish between two cases: $k_{12}\neq 0$, and $k_{12}=0$. In order to describe both cases simultaneously
we proceed as in the previous subsection and replace the third logarithmic term in (\ref{59}) by
\begin{eqnarray}\label{63a}
\left[1-\delta(k_{12})\right]\ln\left[-k_{21}-k_{22}p(0)\mathcal{S}^{-}(0,z)+k_{11}p(1)\mathcal{S}^{+}(1,z)
  +k_{12}p(1)p(0)\mathcal{S}^{-}(0,z)\mathcal{S}^{+}(1,z)\right]\nonumber\\
  +\delta(k_{12})\ln\left[-k_{21}-k_{22}p(0)\mathcal{S}^{-}(0,z)+k_{11}p(1)\mathcal{S}^{+}(1,z)\right]\;.
\end{eqnarray}

For the first terms in (\ref{63a}) the results obtained in (\ref{60}) and (\ref{62}) lead to the expression
\begin{eqnarray}\label{64}
  &&\ln\left[-k_{21}-k_{22}p(0)\mathcal{S}^{-}(0,z)+k_{11}p(1)\mathcal{S}^{+}(1,z)
  +k_{12}p(1)p(0)\mathcal{S}^{-}(0,z)\mathcal{S}^{+}(1,z)\right]\nonumber\\
  &=&\ln k_{12}\sqrt{p(0)p(1)}+2\ln z+\sum_{i=1}^{\infty}\frac{\Lambda_{i}}{z^{i}}\;,
\end{eqnarray}
where the coefficients $\Lambda_{i}$ are obtained through the cumulant expansion
\begin{equation}\label{65}
  \ln\left[1+\sum_{k=1}^{\infty}\frac{\Psi_{k}}{z^{k}}\right]\simeq \sum_{i=1}^{\infty}\frac{\Lambda_{i}}{z^{i}}\;,
\end{equation}
with the definitions
\begin{equation}\label{66}
  \Psi_{1}=\bar{S}_{1}(1)-\bar{S}_{1}(0)-\frac{k_{22}\sqrt{p(0)}+k_{11}\sqrt{p(1)}}{k_{12}\sqrt{p(0)p(1)}}\;,
\end{equation}
and, when $i\geq 2$,
\begin{equation}\label{67}
  \Psi_{i}=\sum_{m=0}^{i}(-1)^{m}\bar{S}_{m}(0)\bar{S}_{i-m}(1)-\frac{\Phi_{i-2}}{k_{12}\sqrt{p(0)p(1)}}\;.
\end{equation}

The asymptotic expansion of the second term in (\ref{63a}) is derived from (\ref{60}) and reads
\begin{eqnarray}\label{68}
  &&\ln\left[-k_{21}-k_{22}p(0)\mathcal{S}^{-}(0,z)+k_{11}p(1)\mathcal{S}^{+}(1,z)
  \right]\nonumber\\
  &=&\ln\left(k_{22}\sqrt{p(0)}+k_{11}\sqrt{p(1)}\right)+\ln z+\sum_{i=1}^{\infty}\frac{\Pi_{i}}{z^{i}}\;,
\end{eqnarray}
where the coefficients $\Pi_{i}$ can be found from the relation
\begin{equation}\label{69}
  \ln\left[1+\frac{1}{k_{22}\sqrt{p(0)}+k_{11}\sqrt{p(1)}}\sum_{k=1}^{\infty}\frac{\Phi_{k-1}}{z^{k}}\right]\simeq\sum_{i=1}^{\infty}\frac{\Pi_{i}}{z^{i}}\;.
\end{equation}

By using (\ref{39}), the expansion (\ref{27}), and the result (\ref{64}) and (\ref{68}) we can conclude that
\begin{eqnarray}\label{70}
  \ln\Delta(iz)&=&-\frac{1}{4}\ln p(0)p(1)+\left[1-\delta(k_{12})\right]\ln k_{12}\sqrt{p(0)p(1)}+\delta(k_{12})\ln\left(k_{22}\sqrt{p(0)}+k_{11}\sqrt{p(1)}\right)\nonumber\\
  &+&\left[2-\delta(k_{12})\right]\ln z-\ln 2z+z\int_{0}^{1}S_{-1}^{+}(t)\diff t+\sum_{i=1}^{\infty}\frac{\mathcal{N}_{i}}{z^{i}}\;,
\end{eqnarray}
where we have replaced the third logarithmic term in (\ref{59}) with the modified expression in (\ref{63a}).
The functions $\mathcal{N}_{i}$ introduced in (\ref{70}) have the expression, for $i=2m+1$, $m\in\mathbb{N}_{0}$,
\begin{equation}\label{71}
  \mathcal{N}_{2m+1}=\int_{0}^{1}S_{2m+1}^{+}(t)\diff t+\left[1-\delta(k_{12})\right]\Lambda_{2m+1}+\delta(k_{12})\Pi_{2m+1}\;,
\end{equation}
and for $i=2m$, $m\in\mathbb{N}^{+}$,
\begin{equation}\label{72}
  \mathcal{N}_{2m}=\int_{0}^{1}S_{2m}^{+}(t)\diff t-D_{2m-1}(0)+\left[1-\delta(k_{12})\right]\Lambda_{2m}+\delta(k_{12})\Pi_{2m}\;.
\end{equation}


The expansions (\ref{48}) and (\ref{70}) represent the only information needed in order
to perform the analytic continuation of the spectral zeta function associated with the
Sturm-Liouville problem (\ref{2}) with separated or coupled boundary conditions.

\section{Analytic Continuation of the Spectral Zeta Function}\label{sec4}

To perform the analytic continuation of the spectral zeta function associated with self-adjoint
Sturm-Liouville problems we will need the results about the asymptotic expansions obtained in the
previous section and the integral
representations (\ref{22}) and (\ref{23}). In the case of separated boundary conditions the
representation (\ref{22}) can be conveniently rewritten as a sum of two terms
\begin{equation}\label{76}
  \zeta^{\mathrm{S}}(s)=\frac{\sin\pi s}{\pi}\int_{0}^{1}\diff z\,z^{-2s}\frac{\partial}{\partial z}\ln\Omega(iz)+
  \frac{\sin\pi s}{\pi}\int_{1}^{\infty}\diff z\,z^{-2s}\frac{\partial}{\partial z}\ln\Omega(iz)\; .
\end{equation}
The first integral represents an analytic function for $\Re(s)<1$ while the second one defines an
analytic function in the region $\Re(s)>1/2$.
In order to analytically continue the spectral zeta function to a region extending to the left of
the strip $1/2<\Re(s)<1$ we need to subtract and add a suitable number of terms of the asymptotic
expansion of $\ln\Omega(iz)$ from the second integral \cite{kirsten01}. By using the first $L+2$
terms of the expansion (\ref{48}) the spectral zeta function can be written as
\begin{equation}\label{77}
 \zeta^{\mathrm{S}}(s)=Z^{\mathrm{S}}(s)+\sum_{i=-1}^{L}A_{i}^{\mathrm{S}}(s)\;.
\end{equation}
The function $Z^{\mathrm{S}}(s)$ is analytic in the region $\Re(s)>-(1+L)/2$ and has the form
\begin{eqnarray}\label{78}
  Z^\mathrm{S}(s)&=&\frac{\sin\pi s}{\pi}\int_{0}^{\infty}\diff z\,z^{-2s}\frac{\partial}{\partial z}\Bigg\{\ln\Omega(iz)-H(z-1)\Bigg[
  -\frac{1}{4}\ln p(0)p(1)+\left[1-\delta(A_{2})\right]\ln A_{2}\sqrt{p(0)}\nonumber\\
  &+&\left[1-\delta(B_{2})\right]\ln B_{2}\sqrt{p(1)}+\delta(A_{2})\ln A_{1}+\delta(B_{2})\ln B_{1}+\left[2-\delta(A_{2})-\delta(B_{2})\right]\ln z\nonumber\\
  &-&\ln 2z+z\int_{0}^{1}S_{-1}^{+}(t)\diff t+\sum_{i=1}^{L}\frac{\mathcal{M}_{i}}{z^{i}}\Bigg]\Bigg\}\;,
\end{eqnarray}
with $H(z-1)$ denoting the Heaviside step function. The functions $A_{i}^{\mathrm{S}}(s)$ are meromorphic for $s\in\mathbb{C}$ and
their expressions are
\begin{equation}\label{80}
  A_{-1}^\mathrm{S}(s)=\frac{\sin\pi s}{\pi}\int_{1}^{\infty}\diff z\,z^{-2s}\frac{\partial}{\partial z}\left[z\int_{0}^{1}S_{-1}^{+}(t)\diff t\right]\;,
\end{equation}
\begin{equation}\label{79}
  A_{0}^\mathrm{S}(s)=\frac{\sin\pi s}{\pi}\int_{1}^{\infty}\diff z\,z^{-2s}\frac{\partial}{\partial z}\left\{\left[2-\delta(A_{2})-\delta(B_{2})\right]\ln z-\ln 2z\right\}\;,
\end{equation}
while, for $i\geq 1$,
\begin{equation}\label{81}
A_{i}^\mathrm{S}(s)=\frac{\sin\pi s}{\pi}\int_{1}^{\infty}\diff z\,z^{-2s}\frac{\partial}{\partial z}\left[\frac{\mathcal{M}_{i}}{z^{i}}\right]\;.
\end{equation}

By using the expression (\ref{77}) and by performing the elementary integrations in (\ref{79})-(\ref{81}) we obtain the following
analytically continued expression for the spectral zeta function
\begin{equation}\label{82}
  \zeta^{\mathrm{S}}(s)=Z^{\mathrm{S}}(s)+\frac{\sin\pi s}{\pi}\left[\frac{1-\delta(A_{2})-\delta(B_{2})}{2s}+\frac{1}{2s-1}\int_{0}^{1}S_{-1}^{+}(t)\diff t-\sum_{i=1}^{L}i\frac{\mathcal{M}_{i}}{2s+i}\right]\;.
\end{equation}
The last expression clearly shows that $\zeta^{\mathrm{S}}(s)$ is a meromorphic function of $s$ with a simple pole at the points $s=1/2$ and $s=-(2k+1)/2$, $k \in \mathbb{N}_0$.


For coupled boundary conditions the spectral zeta function can be analogously written as a sum of two terms
\begin{equation}\label{85}
  \zeta^{\mathrm{C}}(s)=\frac{\sin\pi s}{\pi}\int_{0}^{1}\diff z\,z^{-2s}\frac{\partial}{\partial z}\ln\Delta(iz)+
  \frac{\sin\pi s}{\pi}\int_{1}^{\infty}\diff z\,z^{-2s}\frac{\partial}{\partial z}\ln\Delta(iz)\;,
\end{equation}
where the first integral converges for $\Re(s)<1$ and the second for $\Re(s)>1/2$.
To obtain the analytically continued expression of $\zeta^{\mathrm{C}}(s)$ we proceed as for the case of
separated boundary conditions by subtracting and adding from the second integral in (\ref{85}) $L+2$ terms
of the asymptotic expansion in (\ref{70}). This leads to the expression
\begin{equation}\label{85a}
\zeta^{\mathrm{C}}(s)=Z^{\mathrm{C}}(s)+\sum_{i=-1}^{L}A_{i}^{\mathrm{C}}(s)\;.
\end{equation}
The function $Z^{\mathrm{C}}(s)$ is analytic for $\Re(s)>-(1+L)/2$ and has the expression
\begin{eqnarray}\label{87}
  Z^{\mathrm{C}}(s)&=&\frac{\sin\pi s}{\pi}\int_{0}^{\infty}\diff z\,z^{-2s}\frac{\partial}{\partial z}\Bigg\{\ln\Delta(iz)-H(z-1)\Bigg[
  -\frac{1}{4}\ln p(0)p(1)\nonumber\\
  &+&\left[1-\delta(k_{12})\right]\ln k_{12}\sqrt{p(0)p(1)}
  +\delta(k_{12})\ln\left(k_{22}\sqrt{p(0)}+k_{11}\sqrt{p(1)}\right)\nonumber\\
  &+&\left[2-\delta(k_{12})\right]\ln z-\ln 2z+z\int_{0}^{1}S_{-1}^{+}(t)\diff t+\sum_{i=1}^{L}\frac{\mathcal{N}_{i}}{z^{i}}\Bigg]\Bigg\}\;,
\end{eqnarray}
while the functions $A_{i}^{\mathrm{C}}(s)$ are meromorphic in $s$ and have the form
\begin{equation}\label{87a}
A^{\mathrm{C}}_{-1}(s)=A^{\mathrm{S}}_{-1}(s)\;, \quad A^{\mathrm{C}}_{0}(s)=\frac{\sin\pi s}{\pi}\int_{1}^{\infty}\diff z\,z^{-2s}\frac{\partial}{\partial z}\left\{\left[2-\delta(k_{12})\right]\ln z-\ln 2z\right\}\;,
\end{equation}
and, for $i\geq 1$,
\begin{equation}\label{87b}
A_{i}^\mathrm{C}(s)=\frac{\sin\pi s}{\pi}\int_{1}^{\infty}\diff z\,z^{-2s}\frac{\partial}{\partial z}\left[\frac{\mathcal{N}_{i}}{z^{i}}\right]\;.
\end{equation}
By utilizing (\ref{87a}) and (\ref{87b}) in (\ref{85a}) we get
\begin{equation}\label{86}
  \zeta^{\mathrm{C}}(s)=Z^{\mathrm{C}}(s)+\frac{\sin\pi s}{\pi}\left[\frac{1-\delta(k_{12})}{2s}+\frac{1}{2s-1}\int_{0}^{1}S_{-1}^{+}(t)\diff t-\sum_{i=1}^{L}i\frac{\mathcal{N}_{i}}{2s+i}\right]\;.
\end{equation}

The expressions (\ref{82}) and (\ref{86}) represent the desired analytically continued expressions for the spectral zeta function
and the starting point for the computation of the functional determinant of the Sturm-Liouville operator and of the coefficients of the heat kernel asymptotic expansion.

\section{Presence of Zero Modes}\label{sec5}

The analysis of the spectral zeta function presented in the previous sections is based on the assumption that all eigenvalues are positive, in
particular that no zero modes are present. In fact, this assumption was necessary at the end of Section \ref{sec2}
in order to have a vertical strip in the complex $s$-plane in which the integrals (\ref{22}) and
(\ref{23}) were well defined. When a zero mode is present, however, we need to
consider a modified spectral zeta function in which the eigenvalue $\lambda=0$ is excluded.
This is necessary in order to compute, via spectral zeta function regularization, the functional
determinant of the Sturm-Liouville operator (\ref{1})
with the zero mode extracted. Since $\lambda=0$ is now an eigenvalue we cannot simply deform the
contour of integration to the imaginary axis in the integral representations (\ref{20}) and (\ref{21})
of the spectral zeta function to obtain the expressions (\ref{22}) and (\ref{23}).
In fact, even if we select a suitable contour that only encircles the non-zero eigenvalues,
the subsequent deformation of this contour
to the imaginary axis would encounter a pole at the origin $\lambda=0$.

In order to avoid the appearance of a contribution coming from the pole, we replace the
functions $\Omega(\lambda)$ and $\Delta(\lambda)$ in the
integral representations (\ref{20}) and (\ref{21}) by new functions that vanish at all
non-zero eigenvalues but have a non-vanishing limit as $\lambda\to 0$. This result can
be easily achieved by utilizing in the integral representations (\ref{20}) and
(\ref{21}) the ratio of the functions $\Omega(\lambda)$ and $\Delta(\lambda)$ and their respective
leading behavior as $\lambda\to 0$. This behavior can be found by following the ideas developed in \cite{kirsten04}.

Let $\varphi_{0}(x)$ be the non-trivial solution to the differential equation (\ref{2})
corresponding to $\lambda=0$ and let the inner product of $\varphi_{0}(x)$ and $\varphi_{\lambda}(x)$
in the Hilbert space $\mathscr{L}^{2}(I)$ be
\begin{equation}\label{90}
 \Braket{\varphi_{0}|\varphi_{\lambda}}=\int_{0}^{1}\varphi_{0}^{\ast}(x)\varphi_{\lambda}(x)\diff x\;,
\end{equation}
where the $\ast$ denotes complex conjugation. From equation (\ref{2}) it is not very difficult to prove that
\begin{equation}\label{91}
  \Braket{\varphi_{0}|\mathcal{L}\varphi_{\lambda}}=\lambda^{2}\Braket{\varphi_{0}|\varphi_{\lambda}}\;,
\end{equation}
and integrating by parts the term on the left hand side, we get
\begin{equation}\label{92}
  \left[\varphi_{\lambda}(x)p(x)(\varphi_{0}^{\ast}(x))^{\prime}-\varphi_{0}^{\ast}(x)p(x)\varphi_{\lambda}^{\prime}(x)\right]_{0}^{1}=\lambda^{2}\Braket{\varphi_{0}|\varphi_{\lambda}}\;.
\end{equation}
To find an explicit expression for (\ref{92}) we need to specify the boundary conditions.
For separated boundary conditions, since $\lambda=0$ is an eigenvalue, we have that $\varphi^{\ast}_{0}(x)$
satisfies the boundary conditions at both endpoints
\begin{eqnarray}\label{93}
A_{1}\varphi_{0}^{\ast}(0)+A_{2}p(0)(\varphi_{0}^{\ast})^{\prime}(0)&=&0\;,\label{93a}\\
B_{1}\varphi_{0}^{\ast}(1)+B_{2}p(1)(\varphi_{0}^{\ast})^{\prime}(1)&=&0\;.\label{93b}
\end{eqnarray}
The solution $\varphi_{\lambda}(x)$, instead, is only required to satisfy the initial conditions
\begin{equation}\label{94}
  A_{1}\varphi_{0}(0)+A_{2}p(0)(\varphi_{0})^{\prime}(0)=0\;.
\end{equation}
By using the relations (\ref{9}) in (\ref{93a}) we obtain
\begin{equation}
-p(0)\varphi_{\lambda}^{\prime}(0)\varphi_{0}^{\ast}(0)+\varphi_{\lambda}(0)p(0)(\varphi_{0}^{\ast})^{\prime}(0)=0\;,
\end{equation}
which substituted in (\ref{92}) leads to the result
\begin{equation}\label{95}
  p(1)(\varphi_{0}^{\ast})^{\prime}(1)\varphi_{\lambda}(1)-p(1)\varphi_{0}^{\ast}(1)\varphi_{\lambda}^{\prime}(1)=\lambda^{2}\Braket{\varphi_{0}|\varphi_{\lambda}}\;.
\end{equation}
At this point we use the relation (\ref{93b}) in order to further simplify (\ref{95}). In using (\ref{93b}), however,
we need to distinguish between two cases.
If $B_{1}\neq 0$ we have $\varphi_{0}^{\ast}(1)=-(B_{2}/B_{1})p(1)(\varphi_{0}^{\ast})^{\prime}(1)$ and
substituting this result in (\ref{95}) gives the expression
\begin{equation}\label{96}
  \Omega(\lambda)=\frac{B_{1}}{p(1)(\varphi_{0}^{\ast})^{\prime}(1)}\lambda^{2}\Braket{\varphi_{0}|\varphi_{\lambda}}\;.
\end{equation}
If, instead, $B_{2}\neq 0$ we obtain $p(1)(\varphi_{0}^{\ast})^{\prime}(1)=-(B_{1}/B_{2})\varphi_{0}^{\ast}(1)$ and, from (\ref{95}), the relation
\begin{equation}\label{97}
   \Omega(\lambda)=-\frac{B_{2}}{\varphi_{0}^{\ast}(1)}\lambda^{2}\Braket{\varphi_{0}|\varphi_{\lambda}}\;.
\end{equation}

For coupled boundary conditions the general solution $\varphi_{\lambda}(x)$ to (\ref{2}) can
be written as $\varphi_{\lambda}(x)=\alpha u_{\lambda}(x)+\beta v_{\lambda}(x)$ with $u_{\lambda}(x)$
and $v_{\lambda}(x)$ solutions to (\ref{2}) with the initial conditions (\ref{12}) and (\ref{13}), respectively.
By choosing for the coefficients $\alpha$ and $\beta$ the following values
\begin{equation}\label{98}
  \alpha=v(1)-e^{i\gamma}k_{11}\;,\quad\textrm{and}\quad \beta=-u(1)+e^{i\gamma}k_{12}\;,
\end{equation}
the first equation of (\ref{14}) is automatically satisfied while the left hand side of the second equation becomes the function $\Delta(\lambda)$.
We can therefore conclude that with the choice of $\alpha$ and $\beta$ given by (\ref{98}) the coupled boundary conditions can be written, for $\lambda\neq 0$, as
\begin{equation}\label{99}
  \begin{pmatrix}
\varphi_{\lambda}(1) \\
p(1)\varphi^{\prime}_{\lambda}(1)
\end{pmatrix}=\begin{pmatrix}
0 \\
\Delta(\lambda)
\end{pmatrix}+e^{i\gamma}\mathrm{K}\begin{pmatrix}
\varphi_{\lambda}(0) \\
p(0)\varphi^{\prime}_{\lambda}(0)
\end{pmatrix}\;.
\end{equation}
For $\lambda=0$ we obtain, instead, the boundary conditions (\ref{8})
\begin{equation}\label{100}
  \begin{pmatrix}
\varphi_{0}(1) \\
p(1)\varphi^{\prime}_{0}(1)
\end{pmatrix}=e^{i\gamma}\mathrm{K}\begin{pmatrix}
\varphi_{0}(0) \\
p(0)\varphi^{\prime}_{0}(0)
\end{pmatrix}\;,
\end{equation}
since zero is indeed an eigenvalue. The use of the relations (\ref{99}) and (\ref{100}) in (\ref{92}) lead to the expression
\begin{eqnarray}\label{101}
  \lambda^{2}\Braket{\varphi_{0}|\varphi_{\lambda}}&=&-e^{-i\gamma}\left(k_{11}\varphi_{0}^{\ast}(0)+k_{12}(\varphi_{0}^{\ast})^{\prime}(0)\right)\Delta(\lambda)\nonumber\\
  &+&\left(k_{11}k_{22}-k_{12}k_{21}-1\right)\left(\varphi_{0}^{\ast}(0)\varphi_{\lambda}^{\prime}(0)-(\varphi_{0}^{\ast})^{\prime}(0)\varphi_{\lambda}(0)\right)\;.
\end{eqnarray}
By utilizing (\ref{100}) to prove that $k_{11}\varphi_{0}^{\ast}(0)+k_{12}(\varphi_{0}^{\ast})^{\prime}(0)=\varphi_{0}^{\ast}(1)$ and by recalling that
$\textrm{det}\mathrm{K}=1$, equation (\ref{101}) becomes
\begin{equation}\label{102}
  \Delta(\lambda)=-\lambda^{2}\frac{\Braket{\varphi_{0}|\varphi_{\lambda}}}{\varphi_{0}^{\ast}(1)}\;.
\end{equation}

The relations (\ref{96}), (\ref{97}), and (\ref{102}) clearly show that the leading term of the small-$\lambda$ asymptotic expansion for both $\Omega(\lambda)$ and $\Delta(\lambda)$
is proportional to $\lambda^{2}$. This implies that in order to consider the spectral zeta function with the zero mode extracted, the integral representations (\ref{20}) and (\ref{21})
need to be replaced by
\begin{equation}\label{103}
\zeta_{0}^{\mathrm{S}}(s)=\frac{1}{2\pi i}\int_{\mathcal{C}}\diff\lambda\,\lambda^{-2s}\frac{\partial}{\partial\lambda}\ln\left(\frac{\Omega(\lambda)}{\lambda^{2}}\right)\;,\quad
\textrm{and}\quad \zeta_{0}^{\mathrm{C}}(s)=\frac{1}{2\pi i}\int_{\mathcal{D}}\diff\lambda\,\lambda^{-2s}\frac{\partial}{\partial\lambda}\ln\left(\frac{\Delta(\lambda)}{\lambda^{2}}\right)\;.
\end{equation}
Now, the functions $\lambda^{-2}\Omega(\lambda)$ and $\lambda^{-2}\Delta(\lambda)$ do not vanish as $\lambda\to 0$ and, therefore, we can
shift the contours $\mathcal{C}$ and $\mathcal{D}$ to the imaginary axis to obtain the representations, with $\lambda\to iz$,
\begin{equation}\label{104}
  \zeta_{0}^{\mathrm{S}}(s)=\frac{\sin\pi s}{\pi}\int_{0}^{\infty}\diff z\,z^{-2s}\frac{\partial}{\partial z}\ln\left(\frac{\Omega(iz)}{z^{2}}\right)\;,\quad
\textrm{and}\quad \zeta_{0}^{\mathrm{C}}(s)=\frac{\sin\pi s}{\pi}\int_{0}^{\infty}\diff z\,z^{-2s}\frac{\partial}{\partial z}\ln\left(\frac{\Delta(iz)}{z^{2}}\right)\;.
\end{equation}

Based on the results presented in Section \ref{sec4} we can conclude that the only information needed to perform the analytic continuation of the functions $\zeta_{0}^{\mathrm{S}}(s)$ and $\zeta_{0}^{\mathrm{C}}(s)$ is the asymptotic expansion of $\ln z^{-2}\Omega(iz)$ and $\ln z^{-2}\Delta(iz)$, respectively.
It is clear that these asymptotic expansions can be easily obtained from the ones for
$\ln\Omega(iz)$ and $\ln\Delta(iz)$ computed in Section \ref{sec3} by noticing that
\begin{equation}\label{105}
\ln z^{-2}\Omega(iz)=\ln\Omega(iz)-2\ln z\;,\quad\textrm{and}\quad \ln z^{-2}\Delta(iz)=\ln\Delta(iz)-2\ln z\;.
\end{equation}
This relation implies that the large-$z$ asymptotic expansion of $\ln z^{-2}\Omega(iz)$ and $\ln\Omega(iz)$ are the same except for the coefficient of the term $\ln z$ which in the expansion of $\ln z^{-2}\Omega(iz)$ acquires an additional contribution that equals $-2$. Obviously, a similar statement holds true for the asymptotic expansion of $\ln z^{-2}\Delta(iz)$. The analytic continuation of $\zeta_{0}^{\mathrm{S}}(s)$ and $\zeta_{0}^{\mathrm{C}}(s)$ can be obtained by following the same procedure presented in Section \ref{sec4} and by recalling the relation (\ref{105}). By writing $\zeta_{0}^{\mathrm{S}}(s)$ in (\ref{104}) as a sum of two integrals, as done in (\ref{76}), and by subtracting and adding $L$ terms of the asymptotic expansion of $\ln z^{-2}\Omega(iz)$, which is obtained from (\ref{48}) and the relation (\ref{105}), we find
\begin{equation}\label{106}
\zeta_{0}^{\mathrm{S}}(s)=Z_{0}^{\mathrm{S}}(s)+\frac{\sin\pi s}{\pi}\left[-\frac{1+\delta(A_{2})+\delta(B_{2})}{2s}+\frac{1}{2s-1}\int_{0}^{1}S_{-1}^{+}(t)\diff t-\sum_{i=1}^{L}i\frac{\mathcal{M}_{i}}{2s+i}\right]\;,
\end{equation}
where
\begin{eqnarray}\label{107}
Z_{0}^\mathrm{S}(s)&=&\frac{\sin\pi s}{\pi}\int_{0}^{\infty}\diff z\,z^{-2s}\frac{\partial}{\partial z}\Bigg\{\ln\left(\frac{\Omega(iz)}{z^{2}}\right)-H(z-1)\Bigg[
  -\frac{1}{4}\ln p(0)p(1)+\left[1-\delta(A_{2})\right]\ln A_{2}\sqrt{p(0)}\nonumber\\
  &+&\left[1-\delta(B_{2})\right]\ln B_{2}\sqrt{p(1)}+\delta(A_{2})\ln A_{1}+\delta(B_{2})\ln B_{1}-\left[\delta(A_{2})+\delta(B_{2})\right]\ln z\nonumber\\
  &-&\ln 2z+z\int_{0}^{1}S_{-1}^{+}(t)\diff t+\sum_{i=1}^{L}\frac{\mathcal{M}_{i}}{z^{i}}\Bigg]\Bigg\}\;,
\end{eqnarray}
and $z^{-2}\Omega(iz)$ is obtained from the expression in (\ref{96}) if $B_{1}\neq 0$ and from the expression (\ref{97}) if $B_{2}\neq 0$.

To obtain the analytic continuation of the spectral zeta function $\zeta_{0}^{\mathrm{C}}(s)$ we subtract and add from the integrand in (\ref{104}) the asymptotic expansion of $\ln z^{-2}\Delta(iz)$, obtained from (\ref{70}) and (\ref{105}). By doing so we find that
\begin{equation}\label{110}
  \zeta_{0}^{\mathrm{C}}(s)=Z_{0}^{\mathrm{C}}(s)+\frac{\sin\pi s}{\pi}\left[-\frac{\delta(k_{12})}{2s}+\frac{1}{2s-1}\int_{0}^{1}S_{-1}^{+}(t)\diff t-\sum_{i=1}^{L}i\frac{\mathcal{N}_{i}}{2s+i}\right]\;,
\end{equation}
where $Z_{0}^{\mathrm{C}}(s)$ is an analytic function of $s$ in the region $\Re(s)>-(1+L)/2$ having the expression
\begin{eqnarray}\label{111}
 Z_{0}^{\mathrm{C}}(s)&=&\frac{\sin\pi s}{\pi}\int_{0}^{\infty}\diff z\,z^{-2s}\frac{\partial}{\partial z}\Bigg\{\ln\left(\frac{\Delta(iz)}{z^{2}}\right)-H(z-1)\Bigg[
  -\frac{1}{4}\ln p(0)p(1)\nonumber\\
  &+&\left[1-\delta(k_{12})\right]\ln k_{12}\sqrt{p(0)p(1)}+\delta(k_{12})\ln\left(k_{22}\sqrt{p(0)}+k_{11}\sqrt{p(1)}\right)-\delta(k_{12})\ln z\nonumber\\
  &-&\ln 2z+z\int_{0}^{1}S_{-1}^{+}(t)\diff t+\sum_{i=1}^{L}\frac{\mathcal{N}_{i}}{z^{i}}\Bigg]\Bigg\}\;.
\end{eqnarray}


\section{Functional Determinant of Sturm-Liouville Operators}\label{sec6}

The Sturm-Liouville operator (\ref{1}) is elliptic, self-adjoint and acts on suitable scalar functions defined on a one-dimensional finite interval. For such operators the zeta regularized functional determinant is defined as follows \cite{ray,sarnak87}
\begin{equation}\label{114}
\textrm{det}(\mathcal{L})=\exp\{-\zeta^{\prime}(0)\}\;,
\end{equation}
where the  derivative of the zeta function at the point $s=0$ is computed after $\zeta(s)$ has been analytically continued to a neighborhood of $s=0$. For the case of separated and coupled boundary conditions the analytically continued expressions of $\zeta^{\mathrm{S}}(s)$ and $\zeta^{\mathrm{C}}(s)$ are provided in Section \ref{sec4} and represent the starting point for the computation of the functional determinant.

By differentiating (\ref{82}) with respect to the variable $s$ we obtain
\begin{eqnarray}\label{115}
(\zeta^{\mathrm{S}})^{\prime}(s)&=&(Z^{\mathrm{S}})^{\prime}(s)+\cos(\pi s)\Bigg[\frac{1-\delta(A_{2})-\delta(B_{2})}{2s}+\frac{1}{2s-1}\int_{0}^{1}S_{-1}^{+}(t)\diff t-\sum_{i=1}^{L}i\frac{\mathcal{M}_{i}}{2s+i}\Bigg]\nonumber\\
&-&2\frac{\sin\pi s}{\pi}\left[\frac{1-\delta(A_{2})-\delta(B_{2})}{4s^{2}}+\frac{1}{(2s-1)^{2}}\int_{0}^{1}S_{-1}^{+}(t)\diff t-\sum_{i=1}^{L}i\frac{\mathcal{M}_{i}}{(2s+i)^{2}}\right]\;.
\end{eqnarray}
In the limit $s\to 0$ the last expression simplifies to
\begin{equation}\label{116}
(\zeta^{\mathrm{S}})^{\prime}(0)=(Z^{\mathrm{S}})^{\prime}(0)-\int_{0}^{1}S_{-1}^{+}(t)\diff t-\sum_{i=1}^{L}\mathcal{M}_{i}\;,
\end{equation}
which can be obtained by noting that since the function $Z^{\mathrm{S}}(s)$ is analytic in the region $\Re(s)>-(1+L)/2$, the value $s=0$ can simply be set into the expression for $(Z^{\mathrm{S}})^{\prime}(s)$. The explicit expression for $(Z^{\mathrm{S}})^{\prime}(0)$ is found by differentiating (\ref{78}) and by setting $s=0$, namely
\begin{eqnarray}\label{117}
(Z^{\mathrm{S}})^{\prime}(0)&=&-\ln2\Omega(0)-\frac{1}{4}\ln p(0)p(1)+\left[1-\delta(A_{2})\right]\ln A_{2}\sqrt{p(0)}+\left[1-\delta(B_{2})\right]\ln B_{2}\sqrt{p(1)}\nonumber\\
&+&\delta(A_{2})\ln A_{1}+\delta(B_{2})\ln B_{1}+\int_{0}^{1}S_{-1}^{+}(t)\diff t+\sum_{i=1}^{L}\mathcal{M}_{i}\;.
\end{eqnarray}
The substitution of the result (\ref{117}) in the relation (\ref{116}) and the use of the definition (\ref{114}) leads to the following expression for the functional determinant of the Sturm-Liouville operator $\mathcal{L}^{\mathrm{S}}$ endowed with separated boundary conditions
\begin{equation}\label{118}
\textrm{det}\left(\mathcal{L}^{\mathrm{S}}\right)=2[p(0)p(1)]^{\frac{1}{4}}\Omega(0)\frac{\left(A_{2}\sqrt{p(0)}\right)^{\delta(A_{2})-1}\left(B_{2}\sqrt{p(1)}\right)^{\delta(B_{2})-1}}{A_{1}^{\delta(A_{2})}B_{1}^{\delta(B_{2})}}\;.
\end{equation}


In the case of coupled boundary conditions we differentiate the expression in (\ref{86}) with respect to the variable $s$.
Since $Z^{\mathrm{C}}(s)$ is analytic for $\Re(s)>-(1+L)/2$ and the remaining terms are meromorphic functions of $s$ having
no pole at the origin we set $s=0$ in the differentiated expression of (\ref{86}) to obtain
\begin{equation}\label{119}
(\zeta^{\mathrm{C}})^{\prime}(0)=(Z^{\mathrm{C}})^{\prime}(0)-\int_{0}^{1}S_{-1}^{+}(t)\diff t-\sum_{i=1}^{L}\mathcal{N}_{i}\;.
\end{equation}
After differentiating (\ref{87}) and setting $s=0$ it is not very difficult to obtain the following expression
\begin{eqnarray}\label{120}
(Z^{\mathrm{C}})^{\prime}(0)&=&-\ln2\Delta(0)-\frac{1}{4}\ln p(0)p(1)+\left[1-\delta(k_{12})\right]\ln k_{12}\sqrt{p(0)p(1)}\nonumber\\
&+&\delta(k_{12})\ln\left(k_{22}\sqrt{p(0)}+k_{11}\sqrt{p(1)}\right)+\int_{0}^{1}S_{-1}^{+}(t)\diff t+\sum_{i=1}^{L}\mathcal{N}_{i}\;.
\end{eqnarray}
From (\ref{120}), (\ref{119}) and the definition (\ref{114}) we find, for the functional determinant of the Sturm-Liouville operator (\ref{1}) with coupled boundary conditions, the result
\begin{equation}\label{121}
\textrm{det}\left(\mathcal{L}^{\mathrm{C}}\right)=2[p(0)p(1)]^{\frac{1}{4}}\Delta(0)\frac{\left(k_{12}\sqrt{p(0)p(1)}\right)^{\delta(k_{12})-1}}{\left(k_{22}\sqrt{p(0)}+k_{11}\sqrt{p(1)}\right)^{\delta(k_{12})}}\;.
\end{equation}

The terms $\Omega(0)$ and $\Delta(0)$ in the expressions (\ref{118}) and (\ref{121}) for the functional determinant are
obtained by setting $\lambda=0$ in (\ref{10}) and (\ref{18}), respectively.
The function $\varphi_{0}$ which enters in the expression for $\Omega(0)$ represents a non-trivial solution to
the homogeneous differential equation $\cal{L}\varphi_{0}=0$ satisfying the initial conditions (\ref{9}) and can be found
either numerically or analytically. In the case of $\Delta(0)$, instead, a suitable solution to the homogeneous differential equation satisfies the
initial conditions given in (\ref{12}) and (\ref{13}). Also in this case the solution can either be found numerically or analytically.

The above expressions for the functional determinant have been obtained under the assumption that the spectrum of the Sturm-Liouville
operator (\ref{1}) with separated or coupled boundary conditions is strictly positive. When a zero mode is present, however,
it has to be excluded from the definition of the functional determinant. We will denote the functional determinant of the operator
$\mathcal{L}$ with the zero eigenvalue extracted with a prime as follows: $\textrm{det}^{\prime}(\mathcal{L})$. The zeta regularized
definition of $\textrm{det}^{\prime}(\mathcal{L})$ is, in this case,
\begin{equation}\label{123}
\textrm{det}^{\prime}(\mathcal{L})=\exp\{-\zeta_{0}^{\prime}(0)\}\;,
\end{equation}
in which $\zeta_{0}(s)$ denotes the associated spectral zeta function with the zero mode extracted. In the case of separated boundary conditions we differentiate the analytically continued expression for $\zeta^{\mathrm{S}}_{0}(s)$ in (\ref{106}) and take the limit as $s\to 0$ to get
\begin{equation}\label{123a}
(\zeta^{\mathrm{S}}_{0})^{\prime}(0)=(Z^{\mathrm{S}}_{0})^{\prime}(0)-\int_{0}^{1}S_{-1}^{+}(t)\diff t-\sum_{i=1}^{L}\mathcal{M}_{i}\;.
\end{equation}
In the last expression $(Z^{\mathrm{S}}_{0})^{\prime}(0)$ denotes the result of performing the limit as $s\to 0$ of $(Z^{\mathrm{S}}_{0})^{\prime}(s)$,
computed from (\ref{111}), and using the relations (\ref{96}) and (\ref{97}). In more detail, one has
\begin{eqnarray}\label{123b}
(Z^{\mathrm{S}}_{0})^{\prime}(0)&=&-\ln2\Xi\Braket{\varphi_{0}|\varphi_{0}}-\frac{1}{4}\ln p(0)p(1)+\left[1-\delta(A_{2})\right]\ln A_{2}\sqrt{p(0)}\nonumber\\
&+&\left[1-\delta(B_{2})\right]\ln B_{2}\sqrt{p(1)}+\delta(A_{2})\ln A_{1}+\delta(B_{2})\ln B_{1}+\int_{0}^{1}S_{-1}^{+}(t)\diff t+\sum_{i=1}^{L}\mathcal{M}_{i}\;,
\end{eqnarray}
where we have introduced, for convenience, the following notation
\begin{equation}
\Xi=\left\{\begin{array}{ll}
\frac{B_{1}}{p(1)(\varphi_{0}^{\ast})^{\prime}(1)} & \textrm{if}\; B_{1}\neq 0\\
-\frac{B_{2}}{\varphi_{0}^{\ast}(1)} & \textrm{if}\; B_{2}\neq 0
\end{array}\right.\;.
\end{equation}
From (\ref{123a}), (\ref{123b}), and the definition (\ref{123}) we obtain, for the functional determinant, the expression
\begin{equation}\label{124}
\textrm{det}^{\prime}\left(\mathcal{L}^{\mathrm{S}}\right)=2[p(0)p(1)]^{\frac{1}{4}}\Xi\Braket{\varphi_{0}|\varphi_{0}}\frac{\left(A_{2}\sqrt{p(0)}\right)^{\delta(A_{2})-1}\left(B_{2}\sqrt{p(1)}\right)^{\delta(B_{2})-1}}{A_{1}^{\delta(A_{2})}B_{1}^{\delta(B_{2})}}\;.
\end{equation}


For coupled boundary conditions we differentiate (\ref{110}) and take the limit $s\to 0$ to obtain
\begin{equation}\label{124a}
(\zeta^{\mathrm{C}}_{0})^{\prime}(0)=(Z^{\mathrm{C}}_{0})^{\prime}(0)-\int_{0}^{1}S_{-1}^{+}(t)\diff t-\sum_{i=1}^{L}\mathcal{N}_{i}\;,
\end{equation}
where
\begin{eqnarray}\label{124b}
(Z^{\mathrm{C}}_{0})^{\prime}(0)&=&-\ln2\frac{\Braket{\varphi_{0}|\varphi_{0}}}{\varphi_{0}^{\ast}(1)}-\frac{1}{4}\ln p(0)p(1)+\left[1-\delta(k_{12})\right]\ln k_{12}\sqrt{p(0)p(1)}\nonumber\\
&+&\delta(k_{12})\ln\left(k_{22}\sqrt{p(0)}+k_{11}\sqrt{p(1)}\right)+\int_{0}^{1}S_{-1}^{+}(t)\diff t+\sum_{i=1}^{L}\mathcal{N}_{i}\;,
\end{eqnarray}
is obtained by differentiating (\ref{111}) and by taking the limit $s\to 0$ using (\ref{102}).
The last two expressions together with the definition (\ref{123}) provide the following expression for the functional determinant
of the Sturm-Liouville operator for coupled boundary conditions with the zero mode extracted
\begin{equation}\label{126}
\textrm{det}^{\prime}\left(\mathcal{L}^{\mathrm{C}}\right)=2[p(0)p(1)]^{\frac{1}{4}}\frac{\Braket{\varphi_{0}|\varphi_{0}}}{\varphi_{0}^{\ast}(1)}\frac{\left(k_{12}\sqrt{p(0)p(1)}\right)^{\delta(k_{12})-1}}{\left(k_{22}\sqrt{p(0)}+k_{11}\sqrt{p(1)}\right)^{\delta(k_{12})}}\;.
\end{equation}

We would like to stress that in the expressions (\ref{124}) and (\ref{126}) of the functional determinant with the
zero mode extracted, the eigenfunction $\varphi_{0}$, which is a solution to $\cal{L}\varphi_{0}=0$ satisfying (\ref{93a})-(\ref{93b}) in the case of
separated boundary conditions and (\ref{100}) in the case of coupled boundary conditions, can be computed either numerically or analytically.
From the knowledge of $\varphi_{0}$ one can obtain values for $(\varphi_{0}^{\ast})^{\prime}(1)$, $\varphi_{0}^{\ast}(1)$, and
$\Braket{\varphi_{0}|\varphi_{0}}$, the square of the $\mathscr{L}^{2}$-norm of $\varphi_{0}$.

\section{Coefficients of the Heat Kernel Asymptotic Expansion}\label{sec7}

In this section we will use the analytic continuation of the spectral zeta function of the Sturm-Liouville operator (\ref{1}), endowed with either
separated or coupled boundary conditions presented in Section \ref{sec4}, to compute the coefficients of the small-$t$ asymptotic expansion of the
heat kernel trace $\theta(t)=\textrm{Tr}_{\mathscr{L}^{2}}e^{-t\cal{L}}$. For the Sturm-Liouville operator (\ref{1}) the small-$t$ asymptotic
expansion of $\theta(t)$ has the following general form \cite{gilkey95,grainer71,mina53,vassilevich03}
\begin{equation}\label{128}
\theta(t)=\frac{1}{\sqrt{4\pi t}}\sum_{n=0}^{\infty}a_{\frac{n}{2}}t^{\frac{n}{2}}\;,
\end{equation}
where the terms $a_{n/2}$ denote the heat kernel coefficients. Thanks to the relation that exists between the trace of the heat kernel and the
spectral zeta function given by the Mellin transform \cite{seel68-10-288} one can prove, in the one-dimensional case, that the following
relations hold
\begin{equation}\label{129}
a_{\frac{1}{2}-s}=\Gamma(s)\textrm{Res}\,\zeta(s)\;,
\end{equation}
when $s=1/2$ and $s=-(2n+1)/2$ with $n\in\mathbb{N}_{0}$, and furthermore
\begin{equation}\label{130}
a_{\frac{1}{2}+n}=\frac{(-1)^{n}}{n!}\zeta(-n)\;.
\end{equation}
In the expression (\ref{129}) $\textrm{Res}$ denotes the residue of the spectral zeta function.
Since the relations (\ref{129}) and (\ref{130}) express $a_{n/2}$ in terms of either the residue
or the value of the spectral zeta function at a specific point
of the real line, they will be used in this section for the computation of the coefficients of the small-$t$
expansion of $\theta(t)$ as that information can be extracted from the analytically continued expression of $\zeta(s)$ obtained in Section \ref{sec4}.

For separated boundary conditions the analytically continued expression for the spectral zeta function
is given by (\ref{82}). By choosing $L=2n+1$ with $n\in\mathbb{N}_{0}$ in (\ref{82}) the function $Z^{\mathrm{S}}(s)$ becomes analytic for $\Re(s)>-(n+1)$
and, therefore, does not contribute to the reside of $\zeta^{\mathrm{S}}(s)$ at $s=-(2n+1)/2$. In particular from (\ref{82}) we have
\begin{equation}\label{131}
\textrm{Res}\,\zeta^{\mathrm{S}}\left(\frac{1}{2}\right)=\frac{1}{2\pi}\int_{0}^{1}\frac{\diff t}{\sqrt{p(t)}}\;,
\end{equation}
where we have used the expression in (\ref{28}), and for $n\in\mathbb{N}_{0}$,
\begin{equation}\label{132}
\textrm{Res}\,\zeta^{\mathrm{S}}\left(-\frac{2n+1}{2}\right)=\frac{(-1)^{n}}{2\pi}(2n+1)\mathcal{M}_{2n+1}\;.
\end{equation}
In addition, due to the prefactor $\sin(\pi s)/\pi$ in the expression (\ref{78}), we have that $Z^{\mathrm{S}}(-n)=0$. This implies that
\begin{equation}\label{133}
\zeta^{\mathrm{S}}(0)=\frac{1-\delta(A_{2})-\delta(B_{2})}{2}\;,\quad\textrm{and}\quad \zeta^{\mathrm{S}}(-n)=(-1)^{n+1}n\,\cal{M}_{2n}\;, \quad n\in\mathbb{N}^{+}\;.
\end{equation}

By using the results obtained in (\ref{131}) through (\ref{133}) together with the relations (\ref{129}) and (\ref{130}),
we find the following expression for the heat kernel coefficients when separated boundary conditions are imposed
\begin{equation}\label{134}
a^{\mathrm{S}}_{0}=\frac{1}{2\sqrt{\pi}}\int_{0}^{1}\frac{\diff t}{\sqrt{p(t)}}\;,
\end{equation}
\begin{equation}\label{135}
a^{\mathrm{S}}_{n+1}=-\frac{2^{2n}n!}{\sqrt{\pi}(2n)!}\mathcal{M}_{2n+1}\;,
\end{equation}
which has been obtained by utilizing the formula \cite{sriva12}
\begin{equation}\label{136}
\Gamma\left(-\frac{2n+1}{2}\right)=\frac{\sqrt{\pi}(-1)^{n+1}2^{2n+1}n!}{(2n+1)!}\;,\quad n\in\mathbb{N}_{0}\;.
\end{equation}
For the coefficients with half-integer index we find, instead,
\begin{equation}\label{137}
a^{\mathrm{S}}_{\frac{1}{2}}=\frac{1-\delta(A_{2})-\delta(B_{2})}{2}\;,\quad a^{\mathrm{S}}_{\frac{2n+1}{2}}=-\frac{1}{(n-1)!}\mathcal{M}_{2n}\;,\quad n\in\mathbb{N}^{+}\;.
\end{equation}
By using the explicit expressions for $\cal{M}_{i}$ displayed in the Appendix and the relations (\ref{135}) and (\ref{137}), we find
\begin{eqnarray}
a^{\mathrm{S}}_{1}&=&-\frac{1}{32\sqrt{\pi}}\int_{0}^{1}\frac{1}{\sqrt{p(t)}}\left[16V(t)-\frac{p'(t)^{2}}{p(t)}+4p''(t)\right]\diff t\nonumber\\
&-&\frac{1-\delta(A_{2})}{4A_{2}\sqrt{\pi p(0)}}\left(A_{2}p'(0)+4A_{1}\right)+
\frac{1-\delta(B_{2})}{4B_{2}\sqrt{\pi p(1)}}\left(B_{2}p'(1)-4B_{1}\right)\;,
\end{eqnarray}
and
\begin{eqnarray}
a^{\mathrm{S}}_{\frac{3}{2}}&=&\frac{1}{4}\left[V(0)+V(1)\right]-\frac{1}{64}\left[\frac{p'(0)^{2}}{p(0)}+\frac{p'(1)^{2}}{p(1)}\right]+
\frac{1}{16}\left[p''(0)+p''(1)\right]\nonumber\\
&+&\left[1-\delta(A_{2})\right]\left[\frac{A_1^2}{2A_2^2 p(0)}-\frac{V(0)}{2}+\frac{A_1 p'(0)}{4A_2 p(0) }+\frac{p'(0)^2}{16 p(0)}-\frac{p''(0)}{8}\right]\nonumber\\
&+&\left[1-\delta(B_{2})\right]\left[\frac{B_1^2}{2B_2^2 p(1) }-\frac{V(1)}{2}-\frac{B_1 p'(1)}{4 B_2 p(1) }+\frac{p'(1)^2}{16 p(1)}-\frac{p''(1)}{8}\right]\;.
\end{eqnarray}
Note, that in the above formulas and in the following, the understanding is that the terms proportional to
$1-\delta (A_2)$ and $1-\delta (B_2)$ do not contribute when $A_2=0$, respectively $B_2=0$, despite the presence of $A_2$ and $B_2$
in denominators.


These leading heat kernel coefficients can be compared with known results \cite{gilkey95,kirsten01,vassilevich03}. In order to do so,
the operator (\ref{1}) and boundary conditions (\ref{6}) have to be written using geometric invariants. To this end we want to identify
the Sturm-Liouville operator (\ref{1}) with a Laplacian on the interval $[0,1]$. The second derivative term is matched if we consider $g(x) = p^{-1}(x)$ to
be the metric on the interval. With this metric, the Laplacian for scalars on the interval is
$$ \Delta_I = - \sqrt{p(x)} \frac d {dx} \left( \sqrt{ p(x)} \frac d {dx} \right) = - p(x) \frac{d^2} {dx^2} - \frac 1 2 p' (x) \frac d {dx}\;.$$
The fact that the first order derivative term does not match the one in ${\cal L}$,
$$ {\mathcal L} = - p(x) \frac{d^2}{dx^2} - p' (x) \frac d {dx} + V(x)\;,$$
is incorporated by using a connection one-form $\omega (x) = p' (x) / (4 p(x)).$ It is easily seen that a suitable
rewriting of ${\mathcal L}$ then is
$$ {\mathcal L} = - \sqrt{ p(x)} \left( \frac d {dx} + \omega (x)\right) \sqrt{p (x) } \left( \frac d {dx} + \omega (x) \right) - E\;,$$
where $$E=- \left(V(x) + \frac 1 4 p'' (x) - \frac 1 {16} \frac{(p' (x))^2}{p(x)} \right) $$
is the relevant ``{\it invariant potential}" to use in order to write down the heat kernel coefficients.

For Dirichlet boundary conditions, $A_2 = B_2 =0$, this is all that is needed to verify the above coefficients from known results; note, the
Riemannian volume element is $p^{-1/2} (x)$ and Riemann tensor as well as extrinsic curvature vanish on an interval.

For the cases involving the derivatives in (\ref{6}) we have to identify the relevant parameter as it occurs in Robin boundary conditions. In general
this condition is written as $$ \left[\varphi _{;m}(x)  - R_1 \varphi (x)\right]\left|_{x=0} =0\;, 
\quad \quad \left[\varphi _{;m} (x) - R_2 \varphi (x)\right]\right|_{x=1} =0\;,$$
where $\varphi_{;m}$ denotes the covariant derivative with respect to the {\it exterior} normal. More explicitly for our situation,
$$ \left[ - \sqrt{p(0)} \left( \frac d {dx} + \omega (0) \right) - R_1 \right] \varphi (x)\left|_{x=0} =0\;,
\quad \quad  \left[  \sqrt{p(1)} \left( \frac d {dx} + \omega (1) \right) - R_2 \right] \varphi (x) \right|_{x=1} =0\;.$$
This establishes the relations $$A_1 = - \frac 1 4 \frac{p' (0)}{\sqrt{ p(0)}} - R_1\;, \quad A_2= \frac 1 {\sqrt{p(0)}}\;, \quad B_1 = \frac 1 4 \frac{p' (1)} {\sqrt{p(1)}} - R_2\;, \quad B_2 = \frac 1 {\sqrt{p(1)}}\;,$$
and using these identifications, the known results for the leading coefficients are obtained; note, that the standard notation for $R_i$, $i=1,2$, is $S$ \cite{gilkey95,kirsten01,vassilevich03}.

In addition to making sure that the leading coefficients agree with known results one might wonder how our computations do reproduce the
fact that half-integer coefficients only contain boundary contributions. This means that only values of $p(x)$ and $V(x)$ and their derivatives at 
$x=0$ and $x=1$ appear in the expressions for the heat kernel coefficients with half-integer index.
This can be shown as follows: In (\ref{34}) it is seen that the only possible volume contributions are contained in the term $\int_0^1 {\mathcal S}^+ (t,z) dt$, and, there, the even
powers in the large-$z$ expansion contribute to half-integer coefficients. These terms with even powers of $z$ integrate to boundary terms only. To prove this statement, first note that
$$ {\mathcal S}^{\pm} (x,z) = \pm {\mathcal S}_{odd} (x,z) + {\mathcal S}_{even} (x,z)\;,$$
where $${\mathcal S}_{odd} (x,z) = \sum_{i=-1}^\infty \frac{ S_{2i+1}^+ (x)}{z^{2i+1}}\;, \quad {\mathcal S}_{even} = \sum_{i=0}^\infty \frac{ S_{2i}^+ (x)} {z^{2i}}\;.$$
By construction both, ${\mathcal S}^+ (x,z) $ and ${\mathcal S}^- (x,z)$ satisfy the differential equation (\ref{26}), from which one easily obtains
$$ \left( p(x) {\mathcal S}^+ (x,z) \right) ' + p(x) \left( {\mathcal S}^+ (x,z) \right)^2 = \left( p(x) {\mathcal S}^- (x,z) \right) ' + p(x) \left( {\mathcal S}^- (x,z) \right)^2\;,$$
or, in terms of ${\mathcal S}_{odd}$ and ${\mathcal S}_{even}$,
$$ 2 \left( p(x) {\mathcal S} _{odd} (x,z) \right)' + 4 p(x) {\mathcal S}_{odd} (x,z) {\mathcal S}_{even} (x,z) =0\;.$$
This implies \begin{eqnarray} {\mathcal S}_{even} (x,z) = - \frac 1 2 \frac d {dx} \ln \left( p(x) {\mathcal S}_{odd} (x,z)\right)\;,\label{intbound}\end{eqnarray}
from which the wanted assertion follows.

Let us now focus, now, on the case of coupled boundary conditions. In this case, the relevant analytically
continued expression of the spectral zeta function is (\ref{86}). By choosing once again $L=2n+1$, $n\in\mathbb{N}_{0}$,
the function $Z^{\mathrm{C}}(s)$ becomes analytic in the region $\Re(s)>-(n+1)$ and, in addition,
vanishes for $s=-n$. From the last remark, by using the expression (\ref{86}) in (\ref{129}) and (\ref{130}),
we obtain for the coefficients with integer index
\begin{equation}\label{142}
a_{0}^{\mathrm{C}}=a_{0}^{\mathrm{S}}\;,\quad a^{\mathrm{C}}_{n+1}=-\frac{2^{2n}n!}{\sqrt{\pi}(2n)!}\mathcal{N}_{2n+1}\;,
\end{equation}
with $n\in\mathbb{N}_{0}$, and for those with half-integer index
\begin{equation}\label{143}
a_{\frac{1}{2}}^{\mathrm{C}}=\frac{1-\delta(k_{12})}{2}\;,\quad
a^{\mathrm{C}}_{\frac{2n+1}{2}}=-\frac{1}{(n-1)!}\mathcal{N}_{2n}\;,
\end{equation}
with $n\in\mathbb{N}^{+}$. In particular, for $n=0$ and $n=1$ we have
\begin{eqnarray}\label{144}
a^{\mathrm{C}}_{1}&=&-\frac{1}{32\sqrt{\pi}}\int_{0}^{1}\frac{1}{\sqrt{p(t)}}\left[16V(t)-\frac{p'(t)^{2}}{p(t)}+4p''(t)\right]\diff t
+\frac{\delta(k_{12})}{4\sqrt{\pi}}\left[\frac{k_{11}p'(1)-k_{22}p'(0)+4k_{21}}{\sqrt{p(1)} k_{11}+\sqrt{p(0)} k_{22}}\right]\nonumber\\
&+&\frac{1-\delta(k_{12})}{\sqrt{\pi}}\left[\frac{k_{22}\sqrt{p(0)}+k_{11}\sqrt{p(1)}}{k_{12}\sqrt{p(0)p(1)}}-\frac{p'(0)}{4 \sqrt{p(0)}}+\frac{p'(1)}{4 \sqrt{p(1)}}\right]\;,
\end{eqnarray}
and
\begin{eqnarray}\label{145}
a^{\mathrm{C}}_{\frac{3}{2}}&=&\frac{1}{4}\left[V(0)+V(1)\right]-\frac{1}{64}\left[\frac{p'(0)^{2}}{p(0)}+\frac{p'(1)^{2}}{p(1)}\right]+
\frac{1}{16}\left[p''(0)+p''(1)\right]\nonumber\\
&+&\frac{\left[1-\delta(k_{12})\right]}{2}\Bigg\{\left(\frac{k_{22}\sqrt{p(0)}+k_{11}\sqrt{p(1)}}{k_{12}\sqrt{p(0)p(1)}}\right)^{2}
-(V(0)+V(1))+\frac{1}{8}\left(\frac{p'(0)^2}{p(0)}-\frac{p'(1)^2}{p(1)}\right)\nonumber\\
&-&\frac{1}{4}(p''(0)+p''(1))-\frac{2k_{21}}{k_{12}\sqrt{p(0) p(1)}}-\frac{1}{2 k_{12}}\left[\frac{k_{11} p'(0)}{p(0)}-\frac{ k_{22} p'(1)}{p(1)}\right]\Bigg\}\nonumber\\
&+&\frac{\delta(k_{12})}{32}\Bigg\{\left(\frac{k_{11}p'(1)-k_{22}p'(0)+4k_{21}}{\sqrt{p(1)} k_{11}+\sqrt{p(0)} k_{22}}\right)^{2}\nonumber\\
&-&\frac{1}{\sqrt{p(1)} k_{11}+\sqrt{p(0)} k_{22}}
\Bigg(4\sqrt{p(0)}k_{22}\left(4V(0)+p''(0)\right)\nonumber\\
&+&4\sqrt{p(1)}k_{11}\left(4V(1)+p''(1)\right)
-\frac{k_{22}p'(0)^{2}}{\sqrt{p(0)}}
-\frac{k_{11}p'(1)^{2}}{\sqrt{p(1)}}\Bigg)\Bigg\}\;.
\end{eqnarray}

The set of boundary conditions characterized by $k_{12}=0$ contains, as a particular case, periodic boundary conditions.
These are obtained by imposing, in addition to $k_{12}=0$, the constraints $k_{21} =0,$ $k_{11}=k_{22}=1 $ and, for
$n\in\mathbb{N}_{0}$, $\lim_{t\to 0}p^{(n)}(t)=\lim_{t\to 1}p^{(n)}(t)\neq 0$. From the results obtained in this
Section it is not very difficult to verify that for periodic boundary conditions $a^{\mathrm{C}}_{1/2}=a^{\mathrm{C}}_{3/2}=0$.
Indeed, one can show that all half-integer coefficients vanish. This follows from (\ref{59}) by noting that under the given
assumptions we also have ${\mathcal S}^\pm (0,z) = {\mathcal S}^\pm (1,z)$. Substituting the given values of $k_{ij}$ and using
(\ref{intbound}), one obtains
$$\ln \Delta (iz) = \int\limits_0^1 {\mathcal S}_{odd} (t,z)dt\;,$$
which implies that half-integer coefficients vanish as no even powers in $1/z$ occur.

We would like to point out that higher order heat kernel coefficients for each of the cases presented here
may be found from their general formulas (\ref{135}), (\ref{137}), (\ref{142}), and (\ref{143})
with the help of an algebraic computer program.

\section{Conclusions}

In this paper we have presented a detailed analysis of the analytic continuation
of the spectral zeta function associated with regular, one-dimensional, self-adjoint
Sturm-Liouville problems. The explicit analytically continued expression for the
spectral zeta function has been employed to compute the functional determinant
of the Sturm-Liouville operator and the coefficients of the small-$t$ expansion of
the trace of the heat kernel. The results obtained in this work are very general
as they are valid for any self-adjoint boundary condition and have been obtained
by keeping the functions $p(x)$ and $V(x)$ in the Sturm-Liouville operator (\ref{1})
unspecified. The technique used to perform the analytic continuation of the spectral
zeta function is based on the WKB asymptotic expansion of the eigenfunctions of
the regular Sturm-Liouville problem. This investigation has shown that even if
the eigenvalues and the eigenfunctions of a given problem are
not known explicitly one can still perform the analytic continuation
of the associated spectral zeta function by using only asymptotic information which is obtainable, in general,
from a WKB analysis of the problem. It is important to point out that although in this work we have
focused on regular one-dimensional Sturm-Liouville problems the method developed
and used to perform the analytic continuation of $\zeta(s)$ could be applied with only
few technical modifications to more general problems in higher dimensions.

The spectral zeta function for self-adjoint second order differential operators
containing a smooth potential is extremely useful in the analysis of the Casimir effect
for one-dimensional pistons modeled by potentials \cite{beaure13,fucci1,fucci13}.
In fact, it is well known that in the zeta function regularization method the Casimir energy of
a system is obtained by evaluating the corresponding spectral zeta function at the point
$s=-1/2$ \cite{bord09b,elizalde94}. The results for the analytic continuation of $\zeta(s)$ obtained here
can be used to compute the Casimir energy and force for a one-dimensional piston configuration
endowed with separated or coupled boundary conditions. In this situation some numerical work would be
necessary since the value of $\zeta(s)$ at $s=-1/2$ would acquire
contributions from the analytic function $Z(s)$ appearing in the expressions for
the spectral zeta function in Section \ref{sec4}. This function is defined in terms of a real
integral and can only be computed numerically once the function $p(x)$ and the potential $V(x)$ have
been specified. The study just described would extend to more general
self-adjoint boundary conditions the results on the Casimir effect obtained in \cite{beaure13}
which were limited to one-dimensional potential pistons with Dirichlet boundary conditions.
Investigations in this directions would be very valuable as they would provide, for a piston modeled by a given potential,
the behavior of the Casimir force when different boundary conditions are imposed.

As we have mentioned in the Introduction, regular Sturm-Liouville problems arise naturally
when one considers a Laplace operator acting on scalar functions defined, for instance, on warped
product manifolds of the type $I\times_{f}N$, where $I\subset\mathbb{R}$, $f>0$ is a warping function, and
$N$ a smooth Riemannian manifold. The spectral zeta function for the Laplace operator on warped
product manifolds has been studied in detail in \cite{fucci2} where, however, the analysis
was limited to the cases of Dirichlet and Neumann boundary conditions. The analysis performed in this work
can be utilized to extend the results for the spectral zeta function obtained in \cite{fucci2}  to
more general self-adjoint boundary conditions. In particular, the analytic continuation of $\zeta(s)$ for
coupled boundary conditions performed here would be a major component in the study of the spectral zeta function
for Laplace operators on a warped torus.

It would be particularly interesting to develop a method similar to the one presented in
this paper to obtain the analytic continuation of the spectral zeta function for one-dimensional
\emph{singular} Sturm-Liouville problems. Included in the class of singular Sturm-Liouville problems, for example, is
a second order differential operator with a confining potential acting on functions defined on $\mathbb{R}$.
Also, one obtains a singular problem when the function $p(x)$ or the potential $V(x)$ become
unbounded in a neighborhood of the endpoints of the closed interval $I$. Results for this particular problem
could be important for the analysis of the spectral zeta function associated with a Laplace operator on cuspidal manifolds.

\appendix
\section{The Coefficients $\cal{M}_{i}$ and $\cal{N}_{i}$}

In this appendix we will compute the coefficients $\cal{M}_{i}$ and $\cal{N}_{i}$ up to $i=2$. Obviously, higher order coefficients are easily obtained with the help of a simple computer program.
In order to simplify the notation, for any integrable function $f$ on $I$ we use
\begin{equation}
\left[f\right]=\int_{0}^{1}f(t)\diff t\;.
\end{equation}
By using the relations (\ref{51}) and (\ref{52}), the cumulant expansions (\ref{40}), (\ref{47a}), and (\ref{47b}), and the recurrence relation (\ref{30}), one obtains
\begin{eqnarray}\label{app1}
\cal{M}_{1}&=&\left[\frac{V}{2 \sqrt{p}}-\frac{(p')^2}{32 p^{3/2}}+\frac{p''}{8 \sqrt{p}}\right]+\frac{1-\delta(A_{2})}{4A_{2}\sqrt{p(0)}}\left(A_{2}p'(0)+4A_{1}\right)\nonumber\\
&-&\frac{1-\delta(B_{2})}{4B_{2}\sqrt{p(1)}}\left(B_{2}p'(1)-4B_{1}\right)\;,
\end{eqnarray}
and
\begin{eqnarray}\label{app2}
\cal{M}_{2}&=&\left[-\frac{(p')^3}{64 p^2}-\frac{V'}{4}+\frac{p' p''}{32 p}-\frac{1}{16} p^{(3)}\right]-\frac{V(0)}{2}+\frac{p'(0)^2}{32 p(0)}-\frac{p''(0)}{8}\nonumber\\
&+&\left(1-\delta(A_{2})\right)\left(-\frac{A_1^2}{2A_2^2 p(0)}+\frac{V(0)}{2}-\frac{A_1 p'(0)}{4A_2 p(0) }-\frac{p'(0)^2}{16 p(0)}+\frac{p''(0)}{8}\right)\nonumber\\
&+&\left(1-\delta(B_{2})\right)\left(-\frac{B_1^2}{2B_2^2 p(1) }+\frac{V(1)}{2}+\frac{B_1 p'(1)}{4 B_2 p(1) }-\frac{p'(1)^2}{16 p(1)}+\frac{p''(1)}{8}\right)\;.
\end{eqnarray}
For the coefficients $\cal{N}_{i}$ we use the definitions (\ref{71}) and (\ref{72}), and the expansion (\ref{65}) through (\ref{69}) to find
\begin{eqnarray}
\cal{N}_{1}&=&\left[\frac{V}{2 \sqrt{p}}-\frac{(p')^2}{32 p^{3/2}}+\frac{p''}{8 \sqrt{p}}\right]-\delta(k_{12})\left(\frac{k_{11}p'(1)-k_{22}p'(0)+4k_{21}}{4\left(\sqrt{p(1)} k_{11}+\sqrt{p(0)} k_{22}\right)}\right)\nonumber\\
&-&\left(1-\delta(k_{12})\right)\left(\frac{k_{22}\sqrt{p(0)}+k_{11}\sqrt{p(1)}}{k_{12}\sqrt{p(0)p(1)}}-\frac{p'(0)}{4 \sqrt{p(0)}}+\frac{p'(1)}{4 \sqrt{p(1)}}\right)\;,
\end{eqnarray}
and
\begin{eqnarray}
\cal{N}_{2}&=&\left[-\frac{(p')^3}{64 p^2}-\frac{V'}{4}+\frac{p' p''}{32 p}-\frac{1}{16} p^{(3)}\right]-\frac{V(0)}{2}+\frac{p'(0)^2}{32 p(0)}-\frac{p''(0)}{8}\nonumber\\
&-&\frac{\left(1-\delta(k_{12})\right)}{2}\Bigg\{\left(\frac{k_{22}\sqrt{p(0)}+k_{11}\sqrt{p(1)}}{k_{12}\sqrt{p(0)p(1)}}\right)^{2}
-(V(0)+V(1))+\frac{1}{8}\left(\frac{p'(0)^2}{p(0)}+\frac{p'(1)^2}{p(1)}\right)\nonumber\\
&-&\frac{1}{4}(p''(0)+p''(1))-\frac{2k_{21}}{k_{12}\sqrt{p(0) p(1)}}-\frac{k_{11} p'(0)}{2 p(0) k_{12}}+\frac{ k_{22} p'(1)}{2 p(1) k_{12}}\Bigg\}\nonumber\\
&-&\frac{\delta(k_{12})}{32}\Bigg\{\left(\frac{k_{11}p'(1)-k_{22}p'(0)+4k_{21}}{\sqrt{p(1)} k_{11}+\sqrt{p(0)} k_{22}}\right)^{2}\nonumber\\
&-&\frac{1}{\sqrt{p(1)} k_{11}+\sqrt{p(0)} k_{22}}
\Bigg(4\sqrt{p(0)}k_{22}\left(4V(0)+p''(0)\right)\nonumber\\
&+&4\sqrt{p(1)}k_{11}\left(4V(1)+p''(1)\right)
-\frac{k_{22}p'(0)^{2}}{\sqrt{p(0)}}
-\frac{k_{11}p'(1)^{2}}{\sqrt{p(1)}}\Bigg)\Bigg\}\;.
\end{eqnarray}


\end{document}